\documentclass[a4paper, 10pt]{article}
\usepackage{a4wide}
\usepackage{multicol, float}
\usepackage{lineno,hyperref}
\usepackage{verbatim, euscript}
\usepackage{subfigure}
\usepackage{bbm,amssymb}
\usepackage{amsfonts}
\usepackage{amsmath, mathtools}
\usepackage{amsthm, mathabx}
\usepackage{pictexwd}
\usepackage{graphicx}
\usepackage{rotating}
\usepackage{color}
\usepackage{mathrsfs}
\usepackage{mathtools}
\usepackage{xpatch}
\usepackage[title]{appendix}
\usepackage{natbib}
\usepackage[utf8]{inputenc}

\mathtoolsset{showonlyrefs}

\DeclarePairedDelimiter\floor{\lfloor}{\rfloor}

\theoremstyle{plain}
\newtheorem{The}{Theorem}[section]
\newtheorem{Lem}[The]{Lemma}

\newtheorem{Def}[The]{Definition}
\newtheorem{remark}[The]{Remark}

\newtheorem{Prop}[The]{Proposition}

\newcommand{\E}{\mathbb{E}}

\newcommand{\1}{\mathbbm{1}} 
\newcommand{\N}{\mathbb{N}}

\usepackage[utf8]{inputenc}

\newcommand{\nl}{\newline}
\newcommand{\mP}{\mathbb{P}}

\newcommand{\V}{\mathbb{V}}

\newcommand{\tc}{\textcolor{white}{ii}}

\title{Invasion of cooperative parasites in moderately structured host populations}
\author{Vianney Brouard and Cornelia Pokalyuk}
\date{}

\begin{document}

\maketitle

\begin{abstract}
Certain defence mechanisms of phages against the immune system of their bacterial host rely on
cooperation of phages. Motivated by this example we analyse invasion probabilities of cooperative
parasites in moderately structured host populations. We assume that
hosts occupy the vertices of a configuration model and offspring parasites move to neighbouring sites to infect new hosts. Parasites (usually) reproduce only when infecting a host simultaneously and then  generate many offspring. In this regime
we identify and analyse the spatial scale of the population structure at which invasion of parasites
turns from being an unlikely to a highly probable event.
\end{abstract}

\section{Introduction}

We analyse the invasion probability of parasites in moderately structured host populations. The motivation of this study stems from observations of phage populations. Phages are viruses infecting bacteria. The interest in phages has been growing in recent years because of the growing incidence of multi-drug resistant bacteria. As an alternative to antibiotics, in phage therapy the infected host is inoculated with a population of phages to eliminate the pathogenic bacterial population \citet{Thiel2004}.

Bacteria own various mechanisms to defend against phages, one of these is CRISPR-Cas. This mechanism relies on certain complexes of proteins, that are patrolling in the bacterial cell to detect (and subsequently distroy) genetic material of phages (that the bacterial cell or its ancestors encountered previously and stored at the so called CRISPR-locus in the bacterial genome), see \cite{Rath2015Crispr}. Some phages can block these complexes with mechanisms called anti-CRISPR (ACR) which relies essentially on cooperation of ACR-phages \citet{landsberger2018anti}. Indeed, when a CRISPR-resistant bacterium is attacked by a single ACR-phage, the phage often dies, whereas when several phages attack a bacterium simultaneously or subsequently, they have a good chance to replicate \citet{borges2018bacteriophage}, \citet{landsberger2018anti}.\\ 
The models that have been investigated so far to understand the underlying growth dynamics of ACR-phages and CRISPR-resistant bacterial populations are deterministic models that map the behaviour of well-mixed phage and bacterial populations \citet{landsberger2018anti}. In these models one starts with a relatively large phage population, for which simultaneous or rapid subsequent attacks of phages are likely.

Here we consider a phage population that is initially small. In this setting stochastic effects cannot be ignored. We are interested in the probability that the phage population manages to invade the bacterial population, in the sense that a non-trivial proportion of the bacterial population gets infected and subsequently killed by the phages.

We assume that offspring phages attach to neighbouring bacteria. If the bacterial population is well-mixed, 
offspring numbers of phages need to be very large for simultaneous infections of neighbouring bacteria to  be  likely. However, many bacterial populations are spatially structured, e.g. in biofilms, see \citet{TolkerMolin2000}.  In this case bacteria are only adjacent to a relatively small part of the bacterial population and co-infections of bacteria are common even when offspring numbers of phages are moderate. Consequently, invasion of phages should be more likely in spatially structured bacterial populations than in well-mixed populations. 

Population dynamics involving cooperation have been mainly studied from the perspective of a single population that is divided into defectors and cooperators. In these studies one often is interested if cooperators may prevail or coexist with the population of the defectors,  see e.g. \citet{AllenEtAl2017}, \citet{CzupponPfaffelhuber2017}. Here we consider only cooperators. The survival of the population of cooperators is nonetheless  non-trivial, because the capability of the individuals to cooperate depends on the population structure of another population, the host population. 
 
Even though the motivations of this project come from phages, we think that our results might be also relevant for other host-parasite systems. For example it is believed that the infection of cancer cells with oncolytic viruses, that is viruses that attack cancer cells, is more effective, if a cancer cells are hit by several viruses simultaneously, because in this manner the virus can cope better with the (interferon-based) anti-viral response of the host, see \citet{RodriguezBrenesEtAl2017}.
 
In order to put our study into a general context in the following we will consider instead of a population of phages and bacteria a population of cooperative parasites and hosts. Even though viruses (and in particular phages) are not regarded as parasites by biologists we think it is appropriate to call the involved individuals parasites and hosts, because the population dynamics of the phage population is characterised by the fact that phages are only capable to reproduce in their host, the main feature of parasites. 

Spread of parasites or pathogens in finite host populations has been analysed mainly with respect to epidemiological models, in which only the host population is modeled. Hosts are either susceptible, infected or recovered and the host population is placed on the complete graph or the configuration model, see \citet{BarbourReinert2013}, \citet{BrittonPardoux2019} or \citet{BernsteinEtAl2021}. Here we consider both populations.

We model the spatial structure of the host population by placing hosts on the vertices of a random graph of size $N$ formed according to the configuration model. 
We assume that each host is neighboured by $d_N$ hosts, where $1\ll  d_N \ll N$, and hosts are placed on vertices of a random graph whose edges are arranged according to the configuration model. Initially a single host gets infected by a parasite and $v_N$ offspring parasites are produced. Thereafter the populations evolve in discrete generations. At the beginning of each generation parasites move randomly to neighbouring hosts. Whenever a host gets attacked by at least two parasites the parasites reproduce. If a host gets infected only by a single parasite, the infection is successful only with some small probability $\rho_N$. At parasite reproduction $v_N$ parasites are generated. We show that at the scale $v_N\sim c \sqrt{d_N}$, for some $c>0$, the number of neighbouring hosts that is attacked simultaneously by offspring parasites  is approximately Poisson distributed with parameter $c^2/2$. Furthermore, in the regime $v_N \rho_N \sim x$, for some  $0\leq x \leq 1$ the number of hosts that get successfully infected by single parasites is approximately Poisson distributed as well this time with parameter $x$. (The assumption $x\leq1$ guarantees that invasion due to infections by single parasites is unlikely.) 

We explore the spread of the parasite population within the host population (guided by the analysis of epidemics on random graphs, see \citet{BrittonPardoux2019}, Part III, as well as \citet{BarbourReinert2013}) by couplings with (truncated) Galton-Watson processes (GWP) until $N^\alpha$ hosts get infected for some $\alpha>0$ sufficiently large. In this phase the invasion process is essentially driven by pairs of parasites originating from the same vertex and attacking neighbouring hosts simultaneously as well as parasites attacking hosts alone successfully in the case $\rho_N v_N \rightarrow x$ with $x>0$. Once the number of infected hosts per generation exceeds the level $N^\alpha$, with high probability in a finite number of generations the remaining hosts get infected due to parasites attacking hosts simultaneously from different edges.
Hence, the invasion probability of the parasite population, that is the probability that the host population eventually gets killed, is in the critical scale $v_N \sim c \sqrt{d_N}$  asymptotically equal to  the survival probability of a Galton-Watson process with an offspring distribution that is given by the sum of independent Pois($c^2/2$) and Pois($x$)-distributed random variables.

\section{A host-parasite model with cooperative parasites}
\subsection{Model description and main results}
\label{Model description and main results}
Consider a population of hosts and a population of parasites both located on a random graph. The graph has $N$ vertices and each vertex has $d_{N}$ half-edges. We assume that $d_N N$ is even and half-edges are matched according to the configuration model, i.e. half-edges are paired uniformly at random.

Initially, on each vertex a single host is placed. We start the infection process by infecting a randomly chosen host with a parasite. We say that  parasites infect a host, when the infecting parasites replicate in the host. At replication $v_N$ offspring parasites are generated (independent on the number of infecting parasites) and the host as well as the infecting parasite(s) die(s). 

\begin{figure}[h!]
\centering
\includegraphics[width=0.9 \linewidth]{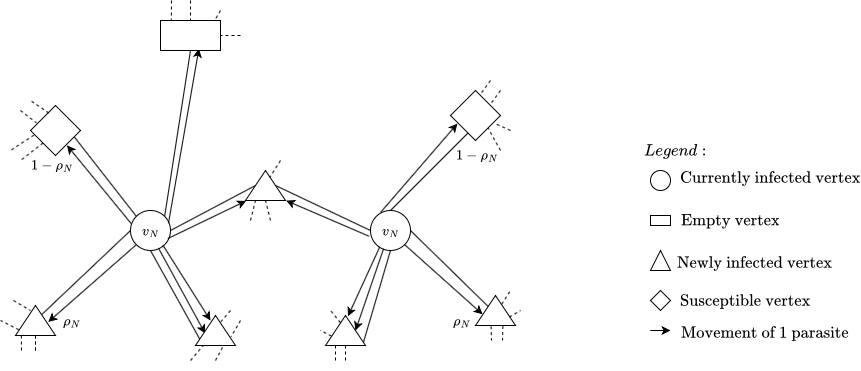}
	\caption{
Illustration of different infection types
	}\label{DiffInfetionTypes}
\end{figure}

The infection process continues in discrete generations according to the following scheme. At the beginning of each generation, parasites move independently to nearest neighbouring vertices. If a vertex to which a parasite moves to is still occupied with a host the parasite attacks this host. If a host is only attacked by a \textit{single parasite}, the parasite replicates only with a small probability $\rho_{N}$. In this case $v_{N}$ offspring parasites are generated and the reproducing parasite as well as the host die. Otherwise (with probability $1-\rho_{N}$), the parasite dies and the host survives. If, however, at least two parasites attack a host simultaneously, the parasites cooperate, they produce (with probability 1) in total $v_{N}$ offspring parasites and the infecting parasites and the host die. 
If a parasite moves to a vertex that is no longer occupied by a host, it stays there and moves further in the next generation. Hosts do not move on the graph during the infection process. See Figure \ref{DiffInfetionTypes} for an illustration of the infection process. 

Given a sequence of parameters $(N, d_N, v_N, \rho_N)_{N\in \mathbb{N}}$ we denote for each $N\in \mathbb{N}$ by 
\begin{linenomath*}
\begin{align*}
 \mathcal{I}^{(N)} = ( I_n^{(N)})_{n\in \mathbb{N}_0},
\end{align*}
\end{linenomath*}
the process that counts the number of infected hosts in the generations $n \in \mathbb{N}_0$ and by 
\begin{linenomath*}
\begin{align*}
\overline{\mathcal{I}}^{(N)} =(\overline{I}_n^{(N)} )_{n\in \mathbb{N}_0} ,
\end{align*}
\end{linenomath*}
with
\begin{linenomath*}
\begin{align*}
    \overline{I}_n^{(N)} := \sum_{i=0}^{n} I_i^{(N)},
\end{align*}
\end{linenomath*}
the process that counts the number of hosts  infected till generation $n\in \mathbb{N}_0$.\\ 
We are interested in the probability that the parasite population invades the host population.
More precisely, we consider the following events.
\begin{Def} {(Invasion of parasites)} \nl 
Consider the above host-parasite model with parameters $(N,d_{N},v_{N}, \rho_{N})_{N \in \mathbb{N}}$. Let $u \in (0,1]$ and denote by 
\begin{linenomath*}
\begin{align*}
E_{u}^{(N)}:= \Big\{ \exists n \in \mathbb{N}_0: \overline{I}^{(N)}_{n}\geq u N \Big\},
\end{align*}
\end{linenomath*}
the event that the parasites invade the host population (at least) to a proportion $u$.
\end{Def}
In the following we consider parameter regimes for which the host population is initially large, that is $N \rightarrow \infty$. We will write $\rightarrow$ for $\overset{N \to \infty}{\rightarrow}$ throughout the paper, unless otherwise specified. We assume that from each host many other hosts can be reached, i.e. $d_N \rightarrow \infty$, but the population is not well mixed, in the sense that $d_N \in o(N)$. Furthermore many offspring parasites are produced at infection of a host, i.e. $v_N \rightarrow \infty$, and the contribution of parasites attacking a host alone is at most critical, in the sense that the expected number of offspring $v_N \rho_N$ generated at such attacks is at most 1. In Theorem \ref{Theorem}  we identify the critical scaling of $v_N$ and $d_N$, at which invasion of a non-trivial proportion of the host population turns from an improbable to a very likely event.
\begin{The}
\label{Theorem}
Assume $d_{N}\in \Theta(N^{\beta})$ 
for some $0<\beta <1$, and $\rho_{N} v_N \rightarrow x$ for some $0\leq x\leq 1$. Depending on the order of $v_{N}$ we obtain the following invasion regimes: \\ 
(i) Assume $v_{N}\in o(\sqrt{d_{N}})$. Then for all $0< u \leq 1$ 
\begin{linenomath*}
\begin{equation}
    \lim_{N \to \infty}\mP \left( E_{u}^{(N)}\right)=0.
\end{equation}
\end{linenomath*}
(ii) Assume $v_{N}\sim c\sqrt{d_{N}}$  for $c>0$. Denote by $\pi(c, x)$ the survival probability of a Galton-Watson process with Pois($\frac{c^{2}}{2} +x$)-offspring distribution. Then the invasion probability of parasites satisfies  for all $0<u\leq 1$
\begin{linenomath*}
    \begin{equation}
     \underset{N \to \infty}{\lim} \mP(E^{(N)}_{u})= \pi(c,x).
    \end{equation}
    \end{linenomath*}
(iii) Assume $\sqrt{d_{N}}\in o(v_{N})$. Then
\begin{linenomath*}
\begin{equation}
 \lim_{N \to \infty}\mP \left( E_{1}^{(N)}\right)=1.
\end{equation}
\end{linenomath*}
\end{The}

After Remark \ref{RemarkNextToTheorem} we will sketch the proof of Theorem \ref{Theorem} in Subsection \ref{Sketch of the proof} and discuss some generalisations of the model and the results in Subsection \ref{Generalisations}. 
A rigorous proof of Theorem \ref{Theorem} will be given in Section \ref{Proof of Main result} after preparing  auxiliary results in Sections \ref{The Upper Galton-Watson Process} - \ref{Death of all bacteria in the graph}. In Table \ref{Table} notation that is frequently used in the manuscript is summarized.  \\
We will often write \textit{whp} for \textit{with high probability} to indicate that an event occurs with a probability that is asymptotically 1 as $N\rightarrow \infty$. 
\begin{remark}\label{RemarkNextToTheorem}
\begin{itemize}
\item[(i)]  In the setting of Theorem \ref{Theorem} (ii) for $\frac{c^{2}}{2} +x \leq 1$ we have $\pi(c,x)=0$, which means that  whp parasites do not invade the host population.
\item[(ii)] We assume $v_N \rho_N \rightarrow x \leq 1$, that is the capability for reproduction of parasites hitting a host alone is subcritical or critical (in the terminology of branching processes).
\item[(iii)] It has been shown that population viscosity, i.e.~limited dispersal of individuals, is generally beneficial for cooperation, see \cite{LionVanBaalen2008}. Here we see an example at which the spatial structure of the host population is passed on to the parasite population that profits from this structure as well. Consequently, in host-parasite systems the host population may on the one hand profit from a spatial structure by enhancing cooperation of hosts, but on the other hand spatial structure may reduce the fitness of the host population because parasite populations may benefit from the spatial structure as well.
\item[(iv)] The proof of Theorem \ref{Theorem} (ii) yields that the time till the entire host population gets infected is upper bounded by $ \frac{(1-\frac{3}{4}\beta + \varepsilon) \log N}{\log (c^2/2+x)}$ for any $\varepsilon>0$, conditioned on a parasite outbreak. Indeed to prove Theorem \ref{Theorem}(ii) we approximate $\mathcal{I}^{(N)}$ by a Galton-Watson process from below, that is truncated from time to time but grows at the same speed as an ordinary Galton-Watson process (with asymptotic offspring mean $c^2/2+x$), until the level $N^{1-\frac{3}{4}\beta + \delta}$ is reached, for some $\delta>0$ sufficiently small. Afterwards the host population gets killed whp within two more generations. From this follows immediately that the host population is whp killed after time $ \frac{(1-\frac{3}{4}\beta + \varepsilon) \log N}{\log (c^2/2+x)}$ for any $\varepsilon >0$ in case of invasion of the parasite population. Similarly, in the setting of Case (iii) it follows directly from the proof (in which couplings between infection processes from Case (iii) and Case (ii) are established, see Section \ref{Proof of Main result} for more details) that the time till extinction of the host population is whp $o(\log(N))$.

With some more effort we expect that it is possible to show that in the setting of Theorem \ref{Theorem}(ii) invasion of the host population ends whp after $\frac{(1-\beta + \varepsilon)\log N}{\log(c^2/2+x)}$ generations. Infection by cooperation of parasites attacking vertices from different edges takes over when the number of infected hosts exceeds the level $N^{1-\beta+ \varepsilon}$, see (the sketch of) the proof of Theorem \ref{Theorem} for more details, subsequently the host population should be killed whp in a finite number of generations.  

Furthermore, depending on the size of the ratio $\frac{v_N^2}{d_N}$ invasion of the host population is considerably faster than $log(N)$ in Case (iii). One shows for example easily that the host population gets whp killed after finitely many generations, if $\frac{v_N^2}{d_N}\sim N^\gamma$ for some $\gamma >0$.
\end{itemize}
\end{remark}

\subsection{Sketch of the proof of Theorem \ref{Theorem}}
\label{Sketch of the proof}
In the following we will use an adaptation of the classical notation for SIR epidemics on a configuration model (see e.g.  \cite{BrittonPardoux2019}, Part III). Define the set of \textit{susceptible hosts} $S_{n}^{(N)}$ as the set of hosts which have not been infected until generation $n$, the set of \textit{infected hosts} $I_{n}^{(N)}$ as the set of hosts which get infected (and killed) at generation $n$, and the set of \textit{removed hosts} $R_{n}^{(N)}$ as the set of hosts which got infected (and killed) strictly before generation $n$. 
Since each host is uniquely related to a vertex, we will sometimes also speak of susceptible vertices and infected vertices instead of susceptible and infected hosts. In addition we will call vertices which hosts have been removed \textit{empty} vertices. 

\begin{figure}[h!]
\centering
\includegraphics[width=0.7\linewidth]{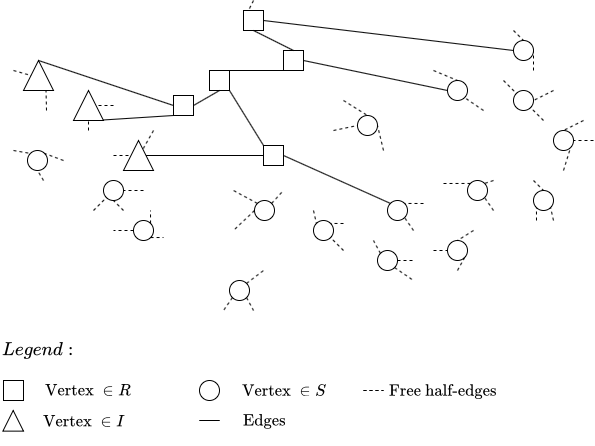}
	\caption{
Illustration of the graph structure
	}\label{ExplorationGraph}
\end{figure}

We explore the random network of hosts while the parasites are spreading in the population. We start at the vertex that got infected initially and build up an edge between two vertices once the edge gets occupied by at least one parasite, see Figure \ref{ExplorationGraph}. 
Half-edges and edges along which parasites move to neighbouring vertices we call \textit{occupied  half-edges} and \textit{occupied edges}, respectively. While an half-edge can get occupied only from a single side (at which it is connected to the vertex), edges can get occupied from two sides. Half-edges and edges that have not been explored yet are called \textit{free half-edges} and \textit{free edges}, respectively. 

We proceed by sketching first the proof in the critical parameter regime $v_N \sim c \sqrt{d_N}$ for some $c>0$, as defined in Theorem \ref{Theorem} (ii). 
In this scaling at the beginning the number of new  infections generated by hosts that got infected in the previous generation is closely related to the birthday problem.  When the number of parasites is relatively small,  offspring parasites from different hosts whp do not interfere and hosts get mainly infected by cooperating parasites that have been generated in the same host and move along the \textit{same} edge, as well as by single parasites attacking successfully neighbouring hosts in the case $x>0$. (In the following we will refer to these single parasites as \textit{successful single parasites}.) Only at a later  stage of the epidemic, when the number of infected and removed hosts exceed the level $N^{1-\beta}$ it gets likely that hosts are infected by parasites that attack the host from \textit{different} edges. Recall that by assumption at parasite reproduction, $v_N$ offspring parasites are generated and a host is connected over $d_N$ half-edges to (roughly) $d_N$ different neighbours. Hence, at the beginning the number of new infections occurring due to cooperation of parasites is for each infected host roughly given by the number of days at which at least two persons share a birthday, when the birthdays of $v_N$ persons are independently and randomly distributed on $d_N$ days.    

If $v_N \sim c \sqrt{d_N}$ for some $c>0$ the number of days at which at least two persons share a birthday is asymptotically $\text{Pois}(\frac{c^2}{2})$-distributed. Furthermore, the number of infections initiated by successful single parasites is asymptotically $\text{Pois}(x)$-distributed, if $x>0$. Since $v_N \rightarrow \infty$, the number of host infections triggered by pairs of parasites moving along the same edge and the number of infections generated by successful single parasites are asymptotically independent. Hence, when the number of infected hosts is still small by each infected host roughly $\text{Pois}( \frac{c^2}{2} + x)$ many new host infections are generated.   \\
Furthermore, offspring parasites of different hosts whp do not interfere at the beginning, hence, for some time the total number of removed  and infected hosts can be estimated from above and below by the total sizes of Galton-Watson-processes with offspring distributions that are close to a $\text{Pois}\left( \frac{c^2}{2}+x\right)$-distribution, see Definition \ref{upper branching process} and \ref{lower branching process} for a rigorous definition of these processes.

To obtain an upper bound on the invasion probability it suffices to prove that whp the total number of removed  and infected hosts can be upper bounded by the total size of the upper Galton-Watson process until a level $\ell_N$ is reached, for some level $\ell_N$ with $\ell_N \rightarrow \infty$. Since the upper Galton-Watson process reaches any level $\ell_N$  with $\ell_N \rightarrow \infty$ with the probability $\pi(c,x) + o(1)$, see Proposition \ref{GWP reaching large size then surviving whp}, the probability to invade the host population up to level $u$ for $0< u\leq 1$ is upper bounded by $\pi(c,x) + o(1)$ as well. 

To derive a lower bound on the invasion probability we couple first $\mathcal{I}^{(N)}$ with a Galton Watson process $\mathcal{Z}_{l}^{(N)}$, such that $\overline{\mathcal{I}}$ is whp bounded from below by the total size of  $\mathcal{Z}_{l}^{(N)}$ until no further hosts are killed or the total number of removed and infected hosts exceeds the threshold $N^\alpha$, for $0<\alpha <\beta$. As for the upper bound, the probability that the total size of the  approximating Galton-Watson process exceeds the threshold $N^\alpha$ is asymptotically equal to $\pi(c,x)$ for any $0<\alpha <\beta$. 

In the case $\beta > \tfrac{4}{7}$ we can choose the level to be reached 
as $N^\alpha$ with $\alpha = 1-\tfrac{3\beta}{4} + \varepsilon$ for some $\varepsilon >0$ small enough such that $1-\tfrac{3\beta}{4} + \varepsilon <\min\left\{\beta, 1- \tfrac{\beta}{2}\right\}$.
Once the level $N^{1-\tfrac{3\beta}{4} + \varepsilon}$ is crossed, whp at most two generations later the total host population gets removed, see Proposition \ref{death all hosts when beta>2/3}. The final epidemic phase is so quick, since once at least $N^{1-\beta}$ hosts are infected, infections generated by pairs of parasites attacking a host from different edges take over. Indeed, if $I_n^{(N)} \in \Theta(N^\gamma)$ in some generation $n$  for some $\gamma>0$ (and $\overline{I}_n^{(N)} \ll N$) $\Theta(v_N N^\gamma)$ offspring parasites are generated. From these parasites $\Theta(N^{2 \gamma + \beta})$ pairs of parasites can be formed. 
 The majority of these pairs consists of parasites that have been generated on different vertices. The probability that such a pair of parasites attacks the same vertex is approximately $\tfrac{1}{N}$. For $\gamma > 1-\beta$ we have $2\gamma + \beta -1 > \gamma$. Hence,
when $\Theta(N^{\gamma})$ hosts are infected for some $1-\beta< \gamma<1 $, more hosts get infected by parasites attacking a vertex from different edges than by pairs of parasites moving along the same edge. Furthermore, for $  1 -\tfrac{3 \beta}{4}< \gamma < 1 -\tfrac{\beta}{2}$  after one generation $\Theta(N^{ 2 \gamma + \beta-1})$ hosts get infected and, since $2 \gamma + \beta -1 > 1- \beta/2$ and $2 (1 -\frac{\beta}{2}) + \beta-1 = 1$, after another generation on average all hosts get killed.    

In the case $\beta \leq \tfrac{4}{7}$ the argument is slightly more involved, since in this case it is not possible to approximate whp $\mathcal{I}^{(N)}$ from below by the Galton-Watson process $\mathcal{Z}_{l}^{(N)}$ until $N^{1-\frac{3\beta}{4} +  \varepsilon}$ hosts get infected. If the number of infected hosts exceeds the level $N^{\beta}$, then with non-trivial probability  an edge is attacked from both ends simultaneously by pairs of parasites or single successful parasites. In this case none of these parasites cause an infection of a host, because the vertices to which these parasites are heading to are already empty. However, we can derive an upper bound on the number of parasites involved in such events and remove the corresponding branches in the lower Galton-Watson process. Since these parasites make up only a vanishing proportion of the total parasite population, the growth of the corresponding truncated Galton-Watson process is asymptotically the same as that of the original Galton-Watson process. Hence, for the truncated Galton-Watson process essentially the same techniques can be applied to finish the proof concerning the probability of invasion in the case $\beta \leq \frac{4}{7}.$

The details of the proof can be found in Sections \ref{The Upper Galton-Watson Process} to  \ref{Proof of Main result}. In Section \ref{The Upper Galton-Watson Process}, we are dealing with an upper bound for the invasion probability. In Section \ref{The Lower Galton-Watson Process} we derive a lower bound of the probability that $N^{\alpha}$ hosts get infected for $0<\alpha \leq 1 - \frac{3}{4}\beta + \varepsilon$. In Section \ref{Death of all bacteria in the graph}, we show that when $N^{1 - \frac{3}{4}\beta + \varepsilon}$ hosts got infected, then whp the remaining hosts will also die in at most two generations. A detailed proof of Theorem \ref{Theorem} (ii) can be found in Section \ref{Proof of Main result}.

In the setting of Theorem \ref{Theorem}(i) the number $v_N$ of offspring parasites generated at an infection is negligible compared to $\sqrt{d_N}$.
Parasites are unlikely to cooperate. Hence, invasion could only be achieved by successful single parasites. But since we are considering the parameters regime $v_{N} \cdot \rho_{N} \rightarrow x\leq 1$, successful single parasites are too rare for invasion. Hence, the parasite population infects only a negligible proportion of the host population before it dies out and so for any $u\in(0,1]$ the invasion probability is $o(1)$. 

On the contrary, if the number $v_N$ of offspring parasites is large compared to $\sqrt{d_N}$, then the infection of a single host leads to an asymptotically infinite number of further host infections. At least one of the infected hosts triggers the invasion of the host population whp.

\subsection{Generalisations}
\label{Generalisations}

The results of Theorem \ref{Theorem} can be extended to more general settings. Next we point out some of these and discuss how the proofs would need to be modified. We carry out detailed proofs only in the setting of Theorem \ref{Theorem} to keep the notation and proofs simple.\\

1.) Instead of assuming that the number $d_N$ of half-edges per vertex and the number $v_N$ of parasite offspring, as well as the probability $\rho_N$ are deterministic, it would also be possible to draw these numbers in an iid manner per vertex/host/parasite according to some distributions $\mathcal{D}^{(N)}$,  $\mathcal{V}^{(N)}$ and $\mathcal{P}^{(N)}$. Our proofs can be easily adapted, if the distributions are sufficiently concentrated. More precisely, this is for example the case, if one can show that for iid random variables $(Y^{(N)}_{i})$ distributed as $\mathcal{D}^{(N)}$, $\mathcal{V}^{(N)}$ with corresponding expectation $\mu_N$ we have that for some $c_N \in o(\mu_N)$  
\begin{linenomath*}
\begin{equation}\label{moreGeneral1}
    \mP \left( \bigcap_{i=1}^{N} \{\lvert Y^{(N)}_{i}-\mu_N \lvert \leq c_N \}\right)=\left(1- \mP \left(\lvert Y^{(N)}_{i}-\mu_N \lvert > c_N \right)\right)^{N} \rightarrow 1, 
\end{equation} 
\end{linenomath*}
and given the total number of parasites, that can be generated, is $M_N$ if the iid random variables $(Y^{(N)}_{i})$ are distributed as $\mathcal{P}^{(N)}$ we have
 \begin{linenomath*}
 \begin{equation}\label{moreGeneral2}
    \mP \left( \bigcap_{i=1}^{M_N} \{\lvert Y^{(N)}_{i}-\mu_N \lvert \leq c_N \}\right)=\left(1- \mP \left(\lvert Y^{(N)}_{i}-\mu_N \lvert > c_N \right)\right)^{M_N} \rightarrow 1. 
\end{equation} 
\end{linenomath*}

This is for $\mathcal{D}^{(N)}$ for example fulfilled if $Y^{(N)}_1$ is distributed as a discretized normal distribution with mean $\mu_N \in \Theta(N^\beta)$ and variance $\sigma_N^2 \in o(N^{2\beta-\delta})$ for some $\delta>0$  or is $Pois(N^{\beta})$. If $(Y^{(N)}_{i})$ has a heavy-tailed distribution with mean $\mu_N = N^\beta$ and $Y^{(N)}_{i} - \mu_N$ has a Pareto-tail, then Condition \eqref{moreGeneral1} is fulfilled, if the tail is of order $\tau >\frac{1}{\beta}$. Similar distributions can be chosen for $\mathcal{V}^{(N)}$ and $\mathcal{P}^{(N)}$.\\

2.) While for many viruses our assumption $v_N \rightarrow \infty$ might be well justified (since viruses often generate a large number of offspring), for some host-parasite systems it might be more appropriate to assume $v_N \equiv v$. If $d_N \rightarrow \infty$, cooperative parasites whp don't invade the host population, as in Theorem \ref{Theorem} (i). If $d_N \equiv d$ (i.e. in a setting of a sparse graph), $v \geq 2$, $d>2$ (for the almost sure existence of a giant component) and $\rho_N v \rightarrow x \in [0,1]$, we expect that some (non zero) proportion of the host population can be infected with some non trivial probability (that asymptotically equals the survival probability of an appropriate Galton-Watson process). \\
After parasite reproduction the $v$ offspring parasites are distributed uniformly at random over the $d$ edges. At the beginning of invasion the parasites that do not occupy the edge, over which the host that generated the offspring parasites got infected, are whp moving to a susceptible vertex. Hence, a suitable candidate for a GWP, which total size approximates the number of infected and removed hosts, should have an offspring distribution that is close to the distribution of the number of the $d-1$ edges that get occupied by at least two parasites or by single successful parasites. One would start the GWP in generation two with a number of lines that equals the random number of hosts that get infected in the first generation.\\
The asymptotic probability to invade a non-trivial proportion of the host population should be equal to the asymptotic survival probability of these GWPes. Given invasion a certain proportion $u, u>0$ of the host population eventually gets infected. The level $u$ should be bounded from below by the survival probability of a suitable approximating backward branching process, see e.g. \cite{BarbourReinert2013} for a construction of such a backward process in the case of a Reed-Frost model. In contrast to the setting of Theorem \ref{Theorem} cooperation from different edges is not sufficiently strong to accelerate the order of the speed at which parasites spread at the end of the invasion process. Indeed, from $\ell_N$ infected hosts by cooperation from different edges of order $(\ell_N )^2/N$ further hosts get infected. This number is of the same order as the number of host that get infected by cooperation over the same edge if  $\ell_N \in \Theta( \ell_N^2/N)$, i.e. only when already of order $N$ hosts are infected.\\ While cooperation from different edges seems not to accelerate the speed of infection, it might lead to the infection of a non-trivial proportion of the host population, since once of order $\Theta(N)$ hosts are infected cooperation from the same edge and cooperation from different edges contribute to the infection process on the same order.\\

3.) In our model we implicitly assume that CRISPR-resistant bacteria get blocked only for a single generation after a phage attack. In reality this blocking may last for a longer time. In this case our result on the asymptotic of the invasion probability remains the same. Indeed, recall that as long as $\overline{I}^{(N)}_n < N^\gamma$ for some $\gamma<\frac{1}{2}(1 -\beta)< 1-\beta$ the number of vertices attacked from different edges is negligible. 
Assuming $\overline{I}^{(N)}_n= N^\gamma$ for some $\gamma < \frac{1}{2}( 1- \beta)$ we also have whp $I^{(N)}_n= \Theta( N^\gamma)$ and the probability that a blocked vertex (which number is of order $N^\gamma v_N$) is attacked by another parasite in generation $n$ is $\mathcal{O}(N^\gamma v_N \tfrac{ N^\gamma v_N}{N})$. This probability is non-trivial for $\gamma\geq \frac{1}{2} (1-\beta)$ (in the setting of Theorem \ref{Theorem}(ii)).
Since invasion of the host population is already decided if the frequency of infected host reaches $N^{\varepsilon}$ for some $\varepsilon >0$, at this stage of the epidemic invasion of the host population occurs anyway with probability $1- o(1).$  \\

4.) In reality the number of offspring parasites generated during an infection could depend on the number of parasites infecting a host.  In the scaling of Theorem \ref{Theorem} (ii)
the probability that a host gets infected by $k$ parasites, for $k\geq 3$, from a set of parasites of size $v_N$ located on the same vertex scales as
$N^{-\frac{(k-2) \beta}{2}}$. As long as $v_N^{(k)} N^{-\frac{(k-2) \beta}{2}} \in o(v_N)$, where $v_N^{(k)}$ is the number of offspring generated at reproduction of $k$ parasites infecting a host, these kind of reproduction events have only a negligible impact on the initial spread of the parasite population. Hence, in this case the asymptotic of the invasion probability remains the same, since parasites generated on different vertices will start to jointly infect hosts only when the frequency of parasites is so high that whp the parasite population will invade the host population anyway.\\

5.) Instead of assuming that the graph on which the epidemic spreads is fixed over the whole time period, one may want to consider evolving graphs, for which edges may be rewired over time. We conjecture that for evolving graphs that rewire at most every generation the results of Theorem \ref{Theorem} remain valid at least if $\beta> \frac{1}{2}$. \\
Indeed the proof of Theorem \ref{Theorem} is to a large extent based on couplings with Galton-Watson processes. For these couplings the number of parasites generated at infection of a host as well as the edges, along which offspring parasites move, are assigned to the vertices independent of the generation when a host gets infected. If the graph is changing over time such a construction could lead to failures of the couplings. However as long as the number of infected hosts of the upper and resp. lower Galton-Watson process coincide exactly with the actual number of infected hosts, this construction yields couplings also for evolving graphs. \\
For the upper bound on the invasion probability we need the coupling to hold until time $\tau_{\ell_N, 0}$ at which the GWP dies out or its total size reaches a level $\ell_N$, for some sequence $\ell_N$ converging to $\infty$ arbitrarily slowly. In the proof of Theorem \ref{Theorem} (ii) we show, that the upper Galton-Watson process and the actual number of infected hosts coincide exactly whp until time $\tau_{\ell_N, 0}$. \\
For the lower bound on the probability of invasion we need to couple the total number of infected hosts with the total size of the lower Galton-Watson process until it reaches the level $N^{1-\beta + \varepsilon}$ for some $\varepsilon >0$ or the GWP dies out. When the level $N^{1-\beta + \varepsilon}$ is reached cooperation from different edges already took over and completes the invasion. The actual number of infected hosts and the number of individuals in the lower Galton-Watson process differs, when vertices get attacked from pairs of parasites originating from different hosts. These events start to play a role when of order $\sqrt{N}$ hosts get infected. If $\beta >\tfrac{1}{2}$,  $N^{1-\beta} \ll \sqrt{N}$, i.e. the lower GWP coincides sufficiently long with $\mathcal{I}^{(N)}$.\\
Similarly, one can adapt the proofs of Theorem \ref{Theorem} (i) and (iii) to the setting of evolving graphs.  In summary, (at least) for $\beta >\tfrac{1}{2}$, the statements of Theorem \ref{Theorem} should also hold for evolving graphs.  \\

6.) Phages that are not able to block CRISPR-resistant bacteria may have a chance to replicate in bacteria that have been blocked by ACR-phages before. However, by a similar reasoning as in item 3.) of this subsection and the sketch of the proof of Theorem \ref{Theorem}(ii) this is only likely when the amount of this type of phages is of order $N^{1-\beta/2}$, that is this type of phages must be much more frequent than ACR-phages initially. \\

7.) In our model we assume that parasites that hit empty vertices keep moving further and hosts are not reproducing. These parasite have only a negligible impact on the fate of the parasite population. Hence, the statements of Theorem \ref{Theorem} remain valid, if we assume that parasites die (or die with a certain probability) when hitting an empty vertex. 

Similarly, if hosts may reproduce (e.g. on empty nearest-neighbour spots) and the offspring numbers per host are sufficiently bounded (e.g. uniformly bounded in $N$) our results remain valid. Indeed, the probability that at least $N^{1-\beta +\varepsilon}$ hosts get infected is asymptotically independent on the state of the vertices on which hosts have been killed already, because the probability to re-hit these vertices is small when the overall number of infected hosts is $\ll N$. After reaching the level $N^{1-\beta +\varepsilon}$ the parasite population expands faster with every generation and in only a finite number of generations the host population gets killed whp. Host reproduction cannot curb this strong parasite expansion, when the offspring numbers are uniformly bounded in $N$.\\

8.) Instead of considering the above configuration models, we could have also considered random $d_N$-regular graphs. For these to exist we would need to assume that $d_N=o(\sqrt{N})$. Furthermore, biologically it seems reasonable that parasites can move from one host to another one over different routes. If several parasites move away from the same vertex this may result in multiple edges, which do not exist for random regular graphs, which makes it more difficult to motivate biologically the consideration of these graphs. Nevertheless given $d_N=o(\sqrt{N})$, we suspect the same result to hold when the configuration model is replaced by the random $d_N$-regular graph model since multiples edges or self loops do not play a role in the infection process.

\section{Upper bound on the invasion probability}
\label{The Upper Galton-Watson Process}
Consider the setting of Theorem \ref{Theorem} (ii). In this section we prepare all results to show that the invasion probability is asymptotically upper bounded by $\pi(c,x)$. We  first introduce the Galton-Watson process $\mathcal{Z}_{u}^{(N)}$, see Definition \ref{upper branching process}. This process is constructed as follows. When the number of infected hosts is sufficiently small and the number of susceptible hosts is still sufficiently large, hosts most likely get infected by pairs of parasites occupying the same half-edge or by successful single parasites. Hence, we estimate the probability that an infected host infects $j$ other hosts, for any $j$ (not too large),  by a lower bound on the sum over $(k, \ell)$ with $k+\ell=j$ of the probabilities that out of $v_N$ parasites, which are originating from the same vertex, $2k$ parasites are distributed as pairs onto $k$ different half-edges, the remaining $v_N - 2k$ parasites are distributed separately on different half-edges and $\ell$ of them are successful single parasites. In all other cases we estimate the number of infected hosts by $v_N$ which is the maximal number of hosts that can get infected by $v_N$ parasites.
We show in Proposition \ref{Coupling from above} that $\overline{\mathcal{I}}^{(N)}$ can whp be estimated from above by the total size of the Galton-Watson process $\mathcal{Z}_{u}^{(N)}$ until it reaches some level $\ell_N$, with $\ell_N \rightarrow \infty$ and $\ell_N \in o(N)$. Only after crossing the level $\ell_N$ it gets likely that two parasites located on different half-edges attack the same host. In this case it could happen that $\overline{\mathcal{I}}^{(N)}$ is no longer dominated by the total size of the Galton-Watson process. However, since the level $\ell_N$ tends to $\infty$, the probability that the total size of $\mathcal{Z}_{u}^{(N)}$ reaches the level $\ell_N$ is asymptotically equal to its survival probability which is asymptotically equal to $\pi(c,x)$, see Proposition \ref{GWP reaching large size then surviving whp}. Consequently, the invasion probability of the host-parasite model is asymptotically bounded from above by $\pi(c,x)$.

\begin{Def}(Upper Galton-Watson process)
\label{upper branching process} \\
Let $0<\delta<\frac{1}{2}$, and $a_{N} \rightarrow \infty$ satisfying $a_N \in o\left( \sqrt{d_N}\right)$. Let $\mathcal{Z}_{u}^{(N)}=\left( Z^{(N)}_{n,u} \right)_{n \in \N_0}$ be a Galton-Watson process with $Z^{(N)}_{0,u}=1$ almost surely, and offspring distribution $\left(p_{j,u}^{(N)}\right)_{j \in \mathbb{N}_{0}}$ with 
\begin{linenomath*}
\begin{align}\label{offspring distribution UGWP 1}
    &p_{j,u}^{(N)}:=\sum_{k+\ell=j}\left(\dfrac{(v_{N}-2a_{N})^{2}}{2d_{N}} \right)^{k}\frac{1}{k!}\exp \left(-\frac{v_{N}^{2}}{2d_{N}}\right)\left(1-\frac{1}{d_N^{\delta}} \right) \dfrac{((v_{N}-2a_{N})\rho_{N})^{\ell}}{\ell!}\left(1-\rho_{N}\right)^{v_{N}}, 
    \end{align}
    \end{linenomath*}
 for all $1\leq  j< a_N$    and
 \begin{linenomath*}
    \begin{align}\label{offspring distribution UGWP 2}
    p_{v_N,u}^{(N)}:=1-\sum_{j=0}^{a_{N}}p^{(N)}_{j,u}.
\end{align}
Denote by $\overline{\mathcal{Z}}_{u}^{(N)} = \left(\overline{Z}^{(N)}_{n,u}\right)_{n\in \mathbbm{N}_0}$ where $\overline{Z}^{(N)}_{n,u} := \sum_{i=0}^n  Z^{(N)}_{i,u}$, that is $\overline{Z}^{(N)}_{n,u}$ gives the total size of $\mathcal{Z}^{(N)}_{u}$ accumulated till generation $n$.
\end{linenomath*}
\end{Def} 
The main results of this section are stated in the next two propositions. 

\begin{Prop}\label{Coupling from above} (Coupling from above) \\ 
Consider a sequence $(\ell_N)_{N \in \N}$ with $\ell_N \rightarrow \infty$ and $\ell_N^3 v_N^2 \in o(N)$. Introduce the stopping time
\begin{linenomath*}
\begin{equation*}
    \begin{split}
        &\tau^{(N)}_{\ell_N, 0} := \inf\left\{ n \in \N_{0}: \overline{Z}^{(N)}_{n,u} \geq  \ell_N \text{ or } Z^{(N)}_{n,u} =0\right\}.
    \end{split}
\end{equation*}
\end{linenomath*}
Then
\begin{linenomath*}
\begin{align*}
    \lim_{N\rightarrow \infty} \mathbb{P}\left( \overline{I}^{(N)}_n \leq  \overline{Z}_{n,u}^{(N)} \ \forall n \leq \tau_{\ell_N, 0}^{(N)}\right) = 1.
\end{align*}
\end{linenomath*}
\end{Prop}

\begin{Prop}
\label{GWP reaching large size then surviving whp}(Probability for the total size of the upper GWP to reach a level $\ell_N$) \\
Consider a sequence $(\ell_N)_{N \in \N}$ with $\ell_N \rightarrow \infty $. Then, we have 
\begin{linenomath*}
\begin{equation*}
    \lim_{N \to \infty}\mP\left(\exists n \in \mathbb{N}_{0}:   \overline{Z}_{n,u}^{(N)} \geq \ell_{N} \right)=\pi(c,x).
\end{equation*}
\end{linenomath*}
\end{Prop}
\tc \\ 
In Subsection \ref{Proof Proposition Coupling above}, we will prove Proposition  \ref{Coupling from above}. In Subsection \ref{General results GWP} we will study (in a  quite general setting) the asymptotic survival probability of a sequence of Galton-Watson processes and afterwards give the proof of Proposition \ref{GWP reaching large size then surviving whp}.

\subsection{Proof of Proposition \ref{Coupling from above}} \label{Proof Proposition Coupling above}

To prepare the proof of Proposition \ref{Coupling from above} we make temporarily two assumptions. 
 First, we ignore infections of hosts by parasites attacking a vertex from different edges. In Proposition \ref{coupling upper GWP} we will show that this assumption is whp fulfilled as long as the number of infected and removed hosts $\overline{I}^{(N)}= R^{(N)}+I^{(N)}$ stays below a certain level $\ell_N$. 
Secondly, we assume that all vertices that get attacked are occupied by hosts and any vertex is connected to exactly $d_N$ different neighbouring vertices. Under the first assumption  this second assumption leads to an upper bound on the number of infected hosts. 

Consider a vertex that is occupied by $v_N$ parasites. 
Denote by $L^{(N)}$ the random number of hosts that get removed after movement of the parasites to neighbouring vertices. \\
The probability distribution of $L^{(N)}$ is given by
\begin{linenomath*}
\begin{equation*}
    \mP(L^{(N)}=0)= \dfrac{d_{N}!}{d_{N}^{v_{N}}(d_{N}-v_{N})!}(1-\rho_{N})^{v_{N}},
    \end{equation*}
    \end{linenomath*}
    and for $k \in \mathbb{N}$
    \begin{linenomath*}
    \begin{align*}
    &\mP(L^{(N)}=k)  =\binom{v_{N}}{k}\dfrac{d_{N}!}{d_{N}^{v_{N}}(d_{N}-v_{N})!}\rho_{N}^{k}(1-\rho_{N})^{v_{N}-k} \\
    & \quad \quad \quad \quad \quad \quad \quad+ \sum_{j=1}^{k} \underset{k_{1}+...+k_{j} \leq v_{N}-(k-j)}{\sum_{k_{1},...,k_{j} \geq 2}} \prod_{\ell=1}^{j}\binom{v_{N}-(k_{1}+...+k_{\ell-1})}{k_{\ell}} \binom{v_{N}-(k_{1}+...+k_{j})}{k-j} \\
    & \quad \quad \quad \quad \quad \quad \quad \quad \cdot \dfrac{1}{\prod_{s=2}^{v_{N}-(k+j)+2}\lvert \{i \in \{1,...,j\}, k_{i}=s \}\rvert !}  \\
    & \quad\quad \quad \quad \quad \quad \quad \quad \cdot \dfrac{d_{N}!}{(d_{N}-j-(v_{N}-(k_{1}+...+k_{j})))!}\rho_{N}^{k-j}(1-\rho_{N})^{v_{N}-(k-j)-(k_{1}+...+k_{j})},
\end{align*}
\end{linenomath*}
because $k$ hosts get infected after movement of $v_N$ parasites if either all parasites move over different edges and exactly $k$ vertices get infected by single successful parasites (and the remaining single parasites are unsuccessful) or if $j$ for $1 \leq j \leq k$ edges get occupied by at least $2$ parasites and the remaining parasites move along different edges and exactly $j-k$ of them are successful.  \\ 
We have $L^{(N)} \leq v_N$ a.s. and, as for the birthday problem, the probability that $L^{(N)}$ is zero is asymptotically 1, if $v_N \in o(\sqrt{d_N})$. In the situation of Theorem \ref{Theorem} (ii), i.e. for $v_N \sim c\sqrt{d_N}$, with $c>0$, the probability that $L^{(N)}$ is zero is asymptotically non-trivial.

Denote by $D^{(N)}_{k,\ell}$ the event that (under the just stated two assumptions) after parasite movement exactly $k+\ell$ hosts get infected by $k$ pairs of parasites moving along the same edge and  $\ell$ successful single parasites, and all the remaining parasites die without infecting a host. The next proposition states that the events $\left(D^{(N)}_{k,\ell}\right)_{k,\ell \in \N_{0}}$ are typical, while all other events occur asymptotically only with negligible probabilities.

\begin{Prop}
\label{only pairs matter at the limit}
Assume the conditions of Theorem \ref{Theorem} (ii) are fulfilled. Then
\begin{linenomath*}
\begin{equation*}
    \lim_{N \to \infty}\mP\left(\bigcup_{k,\ell=0}^{\infty}D^{(N)}_{k,\ell}\right)=1.
\end{equation*}
\end{linenomath*}
\end{Prop}

\begin{proof} 
Let $(k,\ell) \in \N_{0}^{2}$. Denote by 
\begin{linenomath*}
\begin{align}\label{probability w_k}
w_k^{(N)}:= \dfrac{\binom{v_{N}}{2}\binom{v_{N}-2}{2}...\binom{v_{N}-2(k-1)}{2}}{k!d_{N}^{k}}\cdot\dfrac{d_{N}!}{d_{N}^{v_{N}-k}(d_{N}-(v_{N}-k))!} 
\end{align}
\end{linenomath*}
the probability to create exactly $k$ pairs of parasites out of $v_N$ parasites when placing the parasites on $d_N$ spots.
We have
\begin{linenomath*}
\begin{align}
\label{k pairs and alone phages}
    \mP(D^{(N)}_{k,\ell})&=w_k^{(N)}\cdot \binom{v_{N}-2k}{\ell}\rho_{N}^{\ell} \left(1-\rho_{N} \right)^{v_{N}-(2k+\ell)} \\ 
    & \sim \left( \frac{c^{2}}{2}\right)^{k}\frac{1}{k!}\exp\left(-\frac{c^{2}}{2}\right)\cdot\frac{x^{\ell}}{\ell!}\exp \left(-x \right) =: p_{k,\ell}, \notag 
\end{align}
\end{linenomath*}
and for all $j \in \mathbb{N}_0$
\begin{linenomath*}
\begin{equation}
\label{asymptotic poisson (c,x) distribution}
     \mP \left( \bigcup_{k+\ell=j} D^{(N)}_{k,\ell}\right)=\sum_{k+\ell=j} \mP (D^{(N)}_{k,\ell}) \sim \sum_{k+\ell=j}p_{k,\ell}=\left( \frac{c^{2}}{2}+x\right)^{j}\frac{1}{j!}\exp \left(-\left(\frac{c^{2}}{2}+x\right) \right):=p_{j},
\end{equation}
\end{linenomath*}
since the sum of two independent Poisson variables is again Poisson. As the Pois$\left(\frac{c^2}{2}+x\right)$-probability masses $(p_j)_{j\geq 0}$ sum up to 1, we find for all $\varepsilon >0$ a $\widetilde{J}>0$, such that for all $J\geq \widetilde{J}$
\begin{linenomath*}
\begin{align*}1-\varepsilon \leq \sum_{j=0}^{J}p_{j}\leq 1.
\end{align*}
\end{linenomath*}
and by \eqref{asymptotic poisson (c,x) distribution} for $\widetilde{J}$, there exists $\widetilde{N}$ such that for all $N\geq \widetilde{N}$ 
\begin{linenomath*}
\begin{align*}
\bigg\lvert \sum_{j=0}^{\widetilde{J}} \mP\left(\bigcup_{k+\ell=j}D^{(N)}_{k,\ell}\right) -\sum_{k=0}^{\widetilde{J}}p_{j}\bigg\rvert \leq \varepsilon.
\end{align*}
\end{linenomath*}
Consequently
\begin{linenomath*}
\begin{equation*}
     1-2\varepsilon \leq \sum_{j=0}^{\widetilde{J}}\mP\left(\bigcup_{k+\ell=j}D^{(N)}_{k,\ell}\right) \leq \sum_{j=0}^{\infty}\mP\left(\bigcup_{k+\ell=j}D^{(N)}_{k,\ell}\right)\leq 1,
\end{equation*}
\end{linenomath*}
which yields the claim since $\varepsilon$ was arbitrary.
\end{proof}

We show next that the offspring distribution of the upper Galton-Watson process $\mathcal{Z}_{u}^{(N)}$ stochastically dominates $L^{(N)}$ for $N$ large enough, which yields that as long as we can and do ignore infections of hosts by parasites attacking hosts from different edges, $\overline{\mathcal{I}}^{(N)}$ can be upper bounded by $\overline{\mathcal{Z}}_{u}^{(N)}$.  

\begin{Prop}
\label{stochastic majoration pairs}
Under the assumptions of Theorem \ref{Theorem} (ii) the random variables $Z_{1,u}^{(N)}$ and $L^{(N)}$ can be coupled such that for $N$ large enough
\begin{linenomath*}
\begin{align*}
    \mathbb{P}(L^{(N)}\leq Z_{1,u}^{(N)}) =1.
\end{align*}
\end{linenomath*}
\end{Prop}

\begin{proof}
Recall that we denoted by ${(p^{(N)}_{j,u})}_{j \in {\mathbb{N}_0}}$ the offspring distribution of the GWP $\mathcal{Z}_{u}^{(N)}$, see \eqref{offspring distribution UGWP 1} and \eqref{offspring distribution UGWP 2},  and we fixed a level $a_N$ for the definition of $\mathcal{Z}_{u}^{(N)}$. For the proof of the proposition it suffices to show that for $j\leq a_N$ we have $p_{u,j}^{(N)} \leq \mathbbm{P}\left( \bigcup_{k+\ell=j}D_{k,\ell}^{(N)}\right) \leq \mathbbm{P}(L^{(N)} =j)$, since by definition $\mathbbm{P}(Z^{(N)}_{1,u}= v_N) = 1- \mathbbm{P}(Z^{(N)}_{1,u} \leq a_N)$, and $L^{(N)} \leq v_N$ a.s. For all $j \in \mathbb{N}_0$ we have $\mathbbm{P}\left( \bigcup_{k+\ell=j}D_{k,\ell}^{(N)}\right) = \sum_{k+\ell=j} \mathbbm{P}\left( D_{k,\ell}^{(N)}\right) $. We use \eqref{k pairs and alone phages}  and \eqref{probability w_k} to estimate the sum. 
The first factor in Equation \eqref{probability w_k} can be lower bounded by 
$$\dfrac{\binom{v_{N}}{2}\binom{v_{N}-2}{2}\cdots \binom{v_{N}-2(k-1)}{2}}{k!d_{N}^{k}} \geq \left(\dfrac{(v_{N}-2a_{N})^{2}}{2d_{N}} \right)^{k}\frac{1}{k!}, $$
and the second and forth factor of the product in Equation \eqref{k pairs and alone phages} can be lower bounded by  
\begin{linenomath*}
\begin{align*}
    &\binom{v_{N}-2k}{\ell} \geq \frac{(v_{N}-2a_{N})^{\ell}}{\ell!}, \\ 
    &\left(1-\rho_{N} \right)^{v_{N}-(2k+\ell)} \geq \left(1-\rho_{N} \right)^{v_{N}},
\end{align*}
\end{linenomath*}
for $k+\ell \leq a_{N}$.
It remains to estimate the second factor of Equation \eqref{probability w_k}, i.e.  $\dfrac{d_{N}!}{d_{N}^{v_{N}-k}(d_{N}-(v_{N}-k))!}$. Expanding the factorials up to second order we obtain
\begin{linenomath*}
\begin{equation}
\label{asymptotic expansion 1}
    \dfrac{d_{N}!}{d_{N}^{v_{N}-k}(d_{N}-(v_{N}-k))!}=\exp \left(-\frac{(v_{N}-k)^{2}}{2d_{N}}\right) \cdot \left[ 1+\frac{1}{2}\frac{v_{N}-k}{d_{N}}\left(1-\frac{1}{3}\frac{(v_{N}-k)^{2}}{d_{N}}\right)+\mathcal{O}\left( \frac{1}{d_{N}}\right)\right]. 
\end{equation}
\end{linenomath*}
Hence, for $N$ large enough and $0< \delta<\tfrac{1}{2}$
\begin{linenomath*}
\begin{align*}
    \dfrac{d_{N}!}{d_{N}^{v_{N}-k}(d_{N}-(v_{N}-k))!} \geq \exp \left(-\frac{v_{N}^{2}}{2d_{N}}\right)\left(1-\frac{1}{d_{N}^{\delta}} \right),
\end{align*}
\end{linenomath*}
which concludes the proof. 
\end{proof}

So far we ignored infections of hosts by parasites attacking a vertex from different edges. Next we find a sequence of levels $\ell_N$, such that (i) $\ell_N \rightarrow \infty$ and (ii) as long as the number $\overline{I}^{(N)}$ of infected and removed hosts  is bounded by $\ell_N$, these kind of infections are unlikely to occur.

For any $y >0$ denote by
\begin{linenomath*}
\begin{equation*}
    \overline{\tau}_{y}^{(N)}:=\inf \{n \in \N_{0}:  \overline{I}^{(N)}_n \geq y\},
\end{equation*}
\end{linenomath*}
the first time at which the number of infected and removed hosts exceeds the level $y$ and by
\begin{linenomath*}
\begin{equation}
\label{timeCodDiff}
    \tau^{(N)}_{D}:= \inf \{n \in \N_{0}:  \text{ a vertex of } S^{(N)}_{n} \text{ is hit by parasites from different edges} \}. \\ 
\end{equation}
\end{linenomath*}
In the next proposition it is shown that infections of hosts by parasites attacking a vertex from different edges can be neglected as long as the number of infected and removed hosts $\overline{I}^{(N)}$ is of order $o((N/v_N^2)^{\frac{1}{3}})$. 

\begin{Prop}
\label{coupling upper GWP}
Choose a sequence $(\ell_{N})_{N \in \N}$, such that $\ell_N \rightarrow \infty$ and $\ell_N^3v^2_N\in o(N).$ \\ 
Then
\begin{linenomath*}
\begin{equation}\label{noCodiff}
    \lim_{N \to \infty}\mP \left( \tau^{(N)}_{D} \leq \overline{\tau}_{\ell_N}^{(N)}, \overline{\tau}_{\ell_N}^{(N)}<\infty \right)=0.
\end{equation}
\end{linenomath*}
\end{Prop}

\begin{proof}

Recall that we denoted by $S_n^{(N)},I_n^{(N)} $ and $R_n^{(N)}$ the sets of susceptible, infected and empty vertices, resp., in generation $n$.
For the proof of the proposition we need to control the probability that a vertex is hit by at least two parasites from different edges simultaneously.
We first show that it is unlikely to re-hit an already empty vertex till generation $\overline{\tau}^{(N)}_{\ell_N}$. Hence, only parasites on infected vertices remain as candidates for simultaneous infections of parasites from different edges.
However, as we will show below, the number of susceptible vertices till generation $\overline{\tau}^{(N)}_{\ell_N}$ is large and each susceptible vertex has roughly $d_{N}$ free half-edges. That makes it unlikely to hit a susceptible vertex simultaneously from different edges. \\ 
For a rigorous proof denote by $A^{(N)}_n$ the number of parasites on empty vertices in generation $n$ and by 
\begin{linenomath*}
\begin{align*}
\tau_{A}^{(N)}:= \inf \{n \in \mathbb{N}_0 : A_n^{(N)} \geq 1\},    
\end{align*}
\end{linenomath*}
the first generation when at least one parasite hits a vertex of $R^{(N)}$. \\ 
We show next that 
\begin{linenomath*}
\begin{align}
\label{no parasite on recovered vertices}
    \lim_{N \to \infty} \mP(\tau^{(N)}_{A} \leq \overline{\tau}^{(N)}_{\ell_{N}}, \overline{\tau}^{(N)}_{\ell_N} <\infty)=0.
\end{align}
\end{linenomath*}

Let 
\begin{linenomath*}
\begin{align*}
\tau_{\text{no inf}}^{(N)}:=\inf \{n \in \N_0: \ I^{(N)}_n=0\},
\end{align*}
\end{linenomath*}
be the first generation at which no host gets infected. Note that at generation $\tau_{\text{no inf}}^{(N)}$ the infection process is not necessarily finished, as parasites may remain on empty vertices. However, this is whp not the case if $\tau^{(N)}_{\text{no inf}} < \overline{\tau}_{\ell_N}^{(N)}$. More precisely we claim,
\begin{linenomath*}
\begin{equation}
\label{no phage on recovered vertices}
    \lim_{N \to \infty} \mP(\tau^{(N)}_{A} \leq \tau^{(N)}_{\text{no inf}} \wedge \overline{\tau}^{(N)}_{\ell_{N}})=0.
\end{equation}
\end{linenomath*}
 Given we have shown \eqref{no phage on recovered vertices}, we also have \eqref{no parasite on recovered vertices}, since
 \begin{linenomath*}
 \begin{align*}
 \{ \tau^{(N)}_{\text{no inf}} < \tau^{(N)}_A \leq \overline{\tau}^{(N)}_{\ell_{N}}, \overline{\tau}^{(N)}_{\ell_{N}} < \infty\} =\emptyset,
 \end{align*}
 \end{linenomath*}
and hence
\begin{linenomath*}
\begin{align*}
\lim_{N \to \infty} \mP(\tau^{(N)}_{A} \leq \overline{\tau}^{(N)}_{\ell_{N}}, \overline{\tau}^{(N)}_{\ell_N} <\infty)= \lim_{N \to \infty} \mP(\tau^{(N)}_{A} \leq \tau^{(N)}_{\text{no inf}} \wedge \overline{\tau}^{(N)}_{\ell_{N}},\overline{\tau}^{(N)}_{\ell_{N}}<\infty ).
\end{align*}
\end{linenomath*}

So, lets prove \eqref{no phage on recovered vertices}. First of all we have by definition of $\tau^{(N)}_{\text{no inf}}$ that $\tau_{\text{no inf}}^{(N)} \wedge \overline{\tau}_{\ell_N}^{(N)} \leq \ell_N$. Furthermore, the number of parasites generated  in some generation $n$ with $n\leq \tau_{\text{no inf}}^{(N)} \wedge \overline{\tau}_{\ell_N}^{(N)}$ is bounded by $\ell_N v_N$ and
the total number of half-edges formed for vertices of the set $R^{(N)}_{n}$ is at most $\ell_{N}\cdot d_{N}$. The number of half-edges not yet connected to other half-edges in the graph is at least as large as the number of free half-edges of the vertices in the set $S_{n}^{(N)}$, which is bounded from below by $(N-\ell_N)d_{N}-\ell_N v_{N} \geq (N-2 \ell_N)d_{N}$. (Note that the summand $-\ell_{N} v_{N}$ has to be added to account for the potential attacks that do not lead to an infection of a host). Hence, the number of parasites that move to an empty vertex in any generation $n$ with $n \leq \tau_{\text{no inf}}^{(N)} \wedge \overline{\tau}_{\ell_N}^{(N)}$ can be estimated from above by the following iid random variables $(H^{(N)}_n)_{n\in \mathbbm{N}}$. Assume for each $n$ (independently of each other), $\ell_N v_N$ numbers are chosen randomly and without replacement from the set $\{1, ..., (N-2\ell_N) d_N\}$. Let $H_n^{(N)}$ count the numbers falling into the set $\{1, ..., \ell_N d_N\}$.  
Then we have
\begin{linenomath*}
\begin{align*}
    \mP \left(\tau_A^{(N)}  \leq \overline{\tau}_{\ell_N}^{(N)} \wedge \tau_{\text{no inf}}^{(N)}\right) &\leq \mP \left( \exists \ n \leq \ell_N: H^{(N)}_n \geq 1\right) \\
    & \leq \ell_N \mP\left( H^{(N)}_1 \geq 1\right) \\ 
    &\leq \ell_N \left(1- \dfrac{N_1!}{(N_1-l_Nv_{N})!} \cdot
    \dfrac{1}{((N-2\ell_N)d_{N})^{\ell_Nv_{N}}}\right),
\end{align*}
\end{linenomath*}
where $N_1:=(N-3\ell_N)d_{N}$. Using an asymptotic expansion of the factorial, we get 
\begin{linenomath*}
\begin{equation*}
    \dfrac{N_1!}{(N_1-\ell_Nv_{N})!} \cdot
    \dfrac{1}{((N-2l_N)d_{N})^{\ell_Nv_{N}}}=1+\mathcal{O} \left( \frac{\ell_N^{2}v_{N}^2}{N}\right),
\end{equation*}
\end{linenomath*}
so using the assumption $\ell_{N}^{3}v_{N}^{2}=o(N)$, we have proven Equation \eqref{no phage on recovered vertices}. 

To finish the proof of the proposition it remains to show that susceptible vertices are not hit simultaneously by parasites from different edges before generation $\overline{\tau}^{(N)}_{\ell_N}$. Recall the definition of $\tau_{D}^{(N)}$ in \eqref{timeCodDiff}. If $\overline{\tau}_{\ell_N}^{(N)} > \tau_{\text{no inf}}^{(N)}$, then using \eqref{no parasite on recovered vertices} whp $\overline{\tau}_{\ell_N}^{(N)}=\infty$ and hence it suffices to show
\begin{linenomath*}
\begin{equation}
    \lim_{N \to \infty} \mP \left( \tau_D^{(N)} \leq  \overline{\tau}_{\ell_N}^{(N)} \wedge \tau_{\text{no inf}}^{(N)} , \overline{\tau}_{\ell_N}^{(N)} < \infty \right)=0.
\end{equation}
\end{linenomath*}
Denote by $S_{n,{\rm free}}^{(N)}$ the set of susceptible vertices for which all half-edges are still free. As before the number of parasites in the graph is smaller than $\ell_N v_{N}$  for any generation $n$ with $n < \overline{\tau}_{\ell_N}^{(N)} \wedge \tau_{\text{no inf}}^{(N)}$ and $ \lvert S_{n, {\rm free}}^{(N)} \rvert \geq N-\ell_N v_{N} $. Define this time the following sequence of iid random variables $(G^{(N)}_{n})_{n\in \mathbbm{N}}$. 
Consider $N-\ell_N v_N$ boxes each containing $d_N$ balls. Assume (for each $n$ independently) $\ell_N v_N$ balls are drawn randomly and without replacement out of the boxes (that are refilled for each $n$). Let $G^{(N)}_{n}$  be the number of boxes from which at least two balls were drawn. Then we can estimate
\begin{linenomath*}
\begin{align}
\label{NoCoDiffOnSusceptibles}
    \mP \left( \tau_D^{(N)} \leq  \overline{\tau}_{\ell_N}^{(N)} \wedge \tau_{\text{no inf}}^{(N)} , \overline{\tau}_{\ell_N}^{(N)} < \infty \right) &\leq \mP \left( \exists \ n \leq \ell_N: G^{(N)}_{n} \geq 1\right) \notag \\
    & \leq \ell_N \mP\left( G^{(N)}_{1} \geq 1\right) \notag \\ 
    &=\ell_N \left(1-d_{N}^{\ell_{N} v_{N}}\cdot \dfrac{N_2!}{(N_2-\ell_Nv_{N})!} \cdot
    \dfrac{(N_2d_{N}-\ell_Nv_{N})!}{(N_2d_{N})!}\right),
\end{align}
\end{linenomath*}
where $N_2:=N-\ell_N v_{N}$. Using an asymptotic expansion of the factorial, we get 
\begin{linenomath*}
\begin{equation*}
        d_{N}^{\ell_N v_{N}}\cdot \dfrac{N_2!}{(N_2-\ell_Nv_{N})!} \cdot
    \dfrac{(N_2d_{N}-\ell_Nv_{N})!}{(N_2d_{N})!}=1+\mathcal{O}\left(\dfrac{(\ell_Nv_{N})^2}{N} \right),
\end{equation*}
\end{linenomath*}
which shows that the left hand side of \eqref{NoCoDiffOnSusceptibles} converges to 0.
\end{proof}

\begin{proof}[Proof of Proposition \ref{Coupling from above}]
By Proposition \ref{coupling upper GWP} whp no infection of hosts by parasites attacking from different edges occurs till $\overline{\mathcal{I}}^{(N)}$  reaches the level $\ell_N$ for any sequence $(\ell_N)_{N\in \mathbb{N}}$ with $\ell_N \rightarrow\infty$ and $\ell_N^3 v_N^2 \in o(N)$. 
Hence, it suffices to consider the case that such infections do not occur and Proposition \ref{stochastic majoration pairs} can be applied. Consequently, as long as  $\overline{\mathcal{I}}^{(N)}$ has not reached the level $\ell_N$, the number of hosts that get infected from an infected vertex in the next generation can whp be estimated from above by the offspring number of the GWP $\mathcal{Z}_{u}^{(N)}$,  which yields the claim of Proposition \ref{Coupling from above}.
\end{proof}

\subsection{Asymptotic survival probabilities of sequences of Galton-Watson processes and the proof of Proposition 
\ref{GWP reaching large size then surviving whp}
}
\label{General results GWP}
Before we give the proof of Proposition \ref{GWP reaching large size then surviving whp}  we establish some general results about the asymptotic survival probability of a sequence of Galton-Watson processes.

Consider a Galton-Watson process $\mathcal{Z}=\left(Z_n \right)_{n \in \N_0}$ with offspring distribution $\left(p_k\right)_{k \in \mathbb{N}_0}$ and with $Z_0=1$ almost surely, and a sequence of Galton-Watson processes $\mathcal{Z}^{(N)}=\left( Z^{(N)}_{n}\right)_{n \in \N_0}$ with offspring distributions $ \left(p^{(N)}_k\right)_{k \in \mathbb{N}_0}$ and with $Z^{(N)}_0=1$ almost surely, for all $N \in \mathbb{N}$. 

Denote by $\Phi$ and $\Phi^{(N)}$, resp., the probability generating functions of the offspring distributions $\left(p_k\right)_{k \in \mathbb{N}_0}$ and  $\left(p^{(N)}_k\right)_{k \in \mathbb{N}_0}$. Furthermore, denote by $\pi$ and $\pi^{(N)}$ the corresponding survival probabilities, and by $q:=1-\pi$ and $q^{(N)}:=1-\pi^{(N)}$ the corresponding extinction probabilities. Denote also by $\overline{\mathcal{Z}}^{(N)}=\left(\overline{Z}^{(N)}_n:=\sum_{i=0}^{n}Z^{(N)}_i\right)_{n \in \mathbb{N}_0}$ the process that counts the total size of the GWP $\mathcal{Z}^{(N)}$ till generation $n$. \\ 
Recall that $\Phi^{(N)}$ converges uniformly to $\Phi$, if the corresponding offspring distributions converge in total variation distance, in particular, if  
\begin{linenomath*}
\begin{equation}
\label{criterion for uniform convergence}
   \sum_{k=0}^{\infty} \lvert q_{k}^{(N)}-q_{k} \lvert \rightarrow 0, 
\end{equation}
\end{linenomath*}
see \cite{LevinEtAl}, Proposition 4.2, or as one readily checks,
if there exists an $\N-$valued sequence $(K_N)_{N\in \mathbb{N}}$ with $K_N \rightarrow \infty $ such that 
\begin{linenomath*}
\begin{equation}
    \sum_{k=0}^{K_{N}} \lvert q_{k}^{(N)}-q_{k} \lvert \rightarrow 0.
\end{equation}
\end{linenomath*}

\begin{Lem}
\label{Lemma}
Consider the just defined Galton-Watson processes $\mathcal{Z}$ and $(\mathcal{Z}^{(N)})_{N\in \mathbb{N}}$. Furthermore, let $(a_N)_{N\in \mathbb{N}}$ be an $\mathbb{N}$-valued sequence with $a_N \rightarrow \infty$. Assume that the generating functions $(\Phi^{(N)})_{N \in \N}$ converge uniformly in $[0,1]$ to $\Phi$.
Then the following holds:\\
a)
\begin{linenomath*}
\begin{equation*}
    \lvert \pi^{(N)}-\pi \rvert \rightarrow 0,
\end{equation*}
\end{linenomath*}
b)
\begin{linenomath*}
\begin{equation*}
    \mP \left( Z_{a_{N}}^{(N)}=0\right) \rightarrow q,
\end{equation*}
\end{linenomath*}
c) 
\begin{linenomath*}
\begin{equation*}
    \mP \left( \exists n \in \N_0: Z_{n}^{(N)}\geq a_{N} \right) \rightarrow \pi,
\end{equation*}
\end{linenomath*}
d) 
\begin{linenomath*}
\begin{equation*}
    \mP\left(\exists \ n \in \mathbb{N}_0:  \overline{Z}_n^{(N)} \geq a_{N} \right)\rightarrow \pi.
\end{equation*}
\end{linenomath*}
\end{Lem}

\begin{proof}
We show a detailed proof in the case $\pi >0$, with analogous arguments one also shows the claim in the case $\pi=0$. Recall that the extinction probabilities $q$  and $q^{(N)}$ are characterised as the smallest fixed points in $[0,1]$ of the generating functions $\Phi$ and $\Phi^{(N)}$ respectively.
Consider the function
\begin{linenomath*}
\begin{align*}
 g(s) := \Phi(q+s) - (q+s),
\end{align*}
\end{linenomath*}
for $s \in [-q,1-q]$. We have $g(s)=0$, iff $s=0 \text{ or } s=1-q$. Furthermore $g > 0$ for $s < 0$ and $g$ is decreasing up to some $s_0 >0$. \\ 
Let $0 <\varepsilon < s_0$,  
and 
\begin{linenomath*}
\begin{align*}
 \eta < \min\{ g(-\varepsilon) , - g(\varepsilon)\}.
\end{align*}
\end{linenomath*}
Since by assumption $\Phi^{(N)}$ converges uniformly to $\Phi$ 
we find an $N_{0} \in \N$ such that for all $N\geq N_0$ 
\begin{linenomath*}
\begin{align}\label{Eta}
 |\Phi^{(N)}(s) - \Phi(s)| < \eta,
\end{align}
\end{linenomath*}
for all $s\in [0,1]$ and hence
for all $N \geq N_{0}$
\begin{linenomath*}
\begin{align*}
        &\Phi^{(N)}(q -\varepsilon)\geq\Phi(q-\varepsilon)-\eta=g(-\varepsilon)+q-\varepsilon-\eta>q-\varepsilon, \\
        &\Phi^{(N)}(q+\varepsilon)\leq\Phi(q+\varepsilon)+\eta=g(\varepsilon)+q+\varepsilon+\eta<q+\varepsilon.
\end{align*}
\end{linenomath*}
Since $\Phi^{(N)}$ is monotonically increasing on $[0,1]$ and continuous, the smallest non-negative fixed point of $\Phi^{(N)}$ is contained in the interval $[q-\varepsilon, q + \varepsilon]$ which implies a).\\
Denote by $(\Phi\pm\eta)(s):=\Phi(s)\pm \eta$, and $(\Phi\pm\eta)_{n}(s):=(\Phi \pm \eta)\circ \cdots \circ (\Phi \pm \eta)(s)$ the $n$-fold composition of $(\Phi\pm\eta)$. 
An iterated application of \eqref{Eta}
 yields  for all $n\in \mathbb{N}$    
 \begin{linenomath*}\begin{align*}
          \left(\Phi+\eta \right)_n(0)\geq \Phi^{(N)}_n(0) \geq \left(\Phi-\eta \right)_n(0).
\end{align*}
\end{linenomath*}
The sequences $(\left(\Phi-\eta \right)_n(0))_{n\in \mathbbm{N}}$ and $(\left(\Phi +\eta \right)_n(0))_{n\in \mathbbm{N}}$ are increasing and converge for $n\rightarrow \infty$ to the smallest non-negative fixed point of $\Phi-\eta$ and $\Phi +\eta$, respectively. While the fixed point of $\Phi-\eta$ is larger than $q-\varepsilon$, by definition of $\eta$, the fixed point of $\Phi+\eta$ is smaller than $q+\varepsilon$. \\ 
In particular, we have that there exists $\widetilde{n} \in \N$, such that for all  $N \geq N_0$ and for all $ n \geq \widetilde{n}$
\begin{linenomath*}
\begin{align*}
 q-\varepsilon \leq \Phi^{(N)}_{n}(0)\leq q+\varepsilon. 
\end{align*}
\end{linenomath*}
Since $a_N \rightarrow \infty$, there exists $N_1 \in \N$ such that $\forall N \geq N_1, a_N \geq \widetilde{n}$. \\ 
Finally we have for all $N \geq N_2:=\max \{N_0,N_1\}$
\begin{linenomath*}
\begin{equation*}
    q-\varepsilon \leq \Phi^{(N)}_{a_N}(0) \leq q+\varepsilon,
\end{equation*}
\end{linenomath*}
which proves b).\\
The extinction-explosion principle for Galton-Watson processes yields
\begin{linenomath*}
\begin{equation*}
    \mP\left(Z_n^{(N)}>0 \ \forall n \in \N_0\right) \leq  \mP \left( \exists \ n \in \N_0:  Z_n^{(N)} \geq a_N\right).
\end{equation*}
\end{linenomath*}
Hence, by a) 
\begin{linenomath*}
\begin{equation}\label{lower}
    \pi+o(1) \leq \mP \left( \exists \ n \in \N_0:  Z_n^{(N)} \geq a_N\right).
\end{equation}
\end{linenomath*}
Furthermore
\begin{linenomath*}
\begin{align*}
    \pi^{(N)}&=\mP\left(Z_n^{(N)}>0 \ \forall n \in \N_0 \right) \\
    &=\mP\left(\Big{\{} \exists \ n \in \N_0: \ Z_n^{(N)} \geq a_N\Big{\}} \cap \Big{\{} Z_n^{(N)}>0 \ \forall n \in \N_0\Big{\}} \right) \\ 
    &\geq \mP\left( \exists \ n \in \N_0: Z_n^{(N)} \geq a_N\right) \cdot \left(1-\left(q^{(N)}\right)^{a_N} \right).
\end{align*}
\end{linenomath*}
By a) we have that $\mathcal{Z}^{(N)}$ is supercritical for $N$ large enough, which implies
\begin{linenomath*}
\begin{equation*}
    \left(q^{(N)}\right)^{a_{N}} \rightarrow 0.
    \end{equation*}
    \end{linenomath*}
Consequently
\begin{linenomath*}
\begin{align*}
        \mP \left(\exists \ n \in \N_0: \ Z_n^{(N)} \geq a_N\right)& \leq \dfrac{\pi^{(N)}}{1-\left(q^{(N)}\right)^{a_{N}}}  
        =\pi^{(N)}\cdot(1+o(1))  
        =\pi+o(1),
\end{align*}
\end{linenomath*}
which, together with \eqref{lower}, concludes the proof of c).\\
For proving d), it only remains to show that  
\begin{linenomath*}
\begin{equation*}
    \mP \left(\Big{\{}\exists \ n \in \mathbb{N}_0:  \overline{Z}_n^{(N)} \geq a_{N} \Big{\}} \cap \Big{\{} \exists \ n \in \mathbb{N}_0:  Z_n^{(N)}=0\Big{\}}\right)=o(1).
\end{equation*}
\end{linenomath*}
Let $(c_N)_{N\in \mathbbm{N}}$ be a sequence with $c_N \rightarrow \infty$ and $\frac{a_{N}}{c_{N}} \rightarrow \infty$ and consider the subsets
\begin{linenomath*}
\begin{align}
        &A^{(N)}:=\Big{\{}\exists \ n \in \mathbb{N}_0: \overline{Z}_n^{(N)} \geq a_{N}, \ \exists \ i\leq n:  Z_i^{(N)} \geq c_N \Big{\}} \cap \Big{\{} \exists \ n \in \mathbb{N}_0:  Z_n^{(N)}=0\Big{\}}, \\ 
        &B^{(N)}:=\Big{\{}\exists \ n \in \mathbb{N}_0: \overline{Z}_n^{(N)} \geq a_{N}, \ Z_i^{(N)} < c_N \ \forall i\leq n \Big{\}} \cap \Big{\{} \exists \ n \in \mathbb{N}_0:  Z_n^{(N)}=0\Big{\}}.
\end{align}
\end{linenomath*}
By definition
\begin{linenomath*}
\begin{equation*}
        \Big{\{}\exists \ n \in \mathbb{N}_0: \overline{Z}_n^{(N)} \geq a_{N} \Big{\}} \cap \Big{\{} \exists \ n \in \mathbb{N}_0: \  Z_n^{(N)}=0\Big{\}}=  A^{(N)} \sqcup B^{(N)}.
        \end{equation*}
\end{linenomath*}
According to c) we have
\begin{linenomath*}
\begin{equation*}
  \mP \left( A^{(N)}  \right) \leq  \mP \left( \Big{\{} \exists  i \in \N_0: Z_i^{(N)} \geq c_N \Big{\}} \cap \Big{\{} \exists  n \in \mathbb{N}_0:  Z_n^{(N)}=0\Big{\}}\right) \rightarrow 0, 
\end{equation*}
\end{linenomath*}
Furthermore
\begin{linenomath*}
\begin{align*}
        B^{(N)} \subset \Big{\{} Z_{\floor{\frac{a_N}{c_N}}}^{(N)}>0 \Big{\}} \cap \Big{\{} \exists n \in \mathbb{N}_0:  Z_n^{(N)}=0 \Big{\}}, 
\end{align*}
\end{linenomath*}
so according to a) and b) applied with the sequence $\left( \floor{\frac{a_N}{c_N}}\right)_{N\in \mathbbm{N}}$ we get
\begin{linenomath*}
\begin{align*}
    \mP \left({B^{(N)}}^c\right) \geq & \mP\left(\Big{\{} Z_{\floor{\frac{a_N}{c_N}}}^{(N)}= 0 \Big{\}} \sqcup \Big{\{} Z_n^{(N)}>0 \ \forall n \in \mathbb{N}_0 \Big{\}} \right)\\&=\mP \left( \Big{\{} Z_{\floor{\frac{a_N}{c_N}}}^{(N)}=0 \Big{\}}\right)+\mP \left( \Big{\{} Z_n^{(N)}>0 \ \forall n \in \mathbb{N}_0 \Big{\}}\right) \\ 
    &=q+o(1)+\pi+o(1) \\ 
    &=1-o(1),
\end{align*}
\end{linenomath*}
which yields $\mP \left( A^{(N)} \sqcup B^{(N)} \right) \rightarrow 0.$
\end{proof}
We are now ready to prove Proposition \ref{GWP reaching large size then surviving whp}.

\begin{proof}[Proof of Proposition \ref{GWP reaching large size then surviving whp}]
By Lemma \ref{Lemma} d) it suffices to show that the sequence of generating functions $\Phi_{u}^{(N)}$ belonging to the offspring distributions $\left(p_{j,u}^{(N)}\right)_{j \in \N_0}$  of $\mathcal{Z}_{u}^{(N)}$ converges uniformly on $[0,1]$ to the generating function $\Phi^{(c,x)}$ of the $\text{Pois}\left(\frac{c^2}{2}+x\right)$-distribution. We will denote by $(p_{j})_{j \in \N_{0}}$ the probability weights of the $\text{Pois}\left(\frac{c^2}{2}+x\right)$-distribution. According to the remark just before Lemma \ref{Lemma} it suffices to find a sequence $(K_{N})_{N \in \N}$ with $K_{N} \rightarrow \infty$ for which $\sum_{j=0}^{K_{N}} \lvert p^{(N)}_{j,u}-p_{j}\lvert \rightarrow 0.$ We set $K_N = a_N$ and use in the following calculation the asymptotics
\begin{linenomath*}
\begin{align}
     &\left( \dfrac{(v_{N}-2a_{N})^{2}}{2d_{N}}\right)^{k}\exp \left( \dfrac{-v_{N}^{2}}{2d_{N}}\right)=\left( \dfrac{c^{2}}{2}\right)^{k}\exp \left( \dfrac{-c^{2}}{2}\right)\left( 1-h_{N}\right)^{k+1}, \\ 
     &((v_{N}-2a_{N})\rho_{N})^{\ell}(1-\rho_{N})^{v_{N}}=x^{\ell}\exp(-x)(1+o(1))^{\ell+1},
\end{align} 
\end{linenomath*}
where $(h_N)_{N \in \mathbb{N}}$ denotes some appropriate sequence of order  $\mathcal{O} \left(\frac{\max\{ a_{N},r_{N}\}}{\sqrt{d_{N}}} \right)$ and $r_{N}:=v_{N}-c\sqrt{d_{N}}$.
For all $j\geq 0$ 
\begin{linenomath*}
\begin{equation}
     \lvert p^{(N)}_{j,u}-p_{j}\lvert \leq \sum_{k+\ell=j}\frac{1}{k!}\left(\dfrac{c^{2}}{2} \right)^{k}\frac{x^{\ell}}{\ell!}\exp \left(-\left(\dfrac{c^{2}}{2}+x\right) \right) \Big{\lvert} \left(1-h_{N}\right)^{k+1}(1+o(1))^{\ell+1}-1 \Big{\lvert}.
\end{equation}
\end{linenomath*}
The last term can be upper bounded in the following way
\begin{linenomath*}
\begin{align*}
    \vert \left(1-h_{N}\right)^{k+1}(1+o(1))^{\ell+1}-1 \vert &\leq
    h_{N}\Big{\vert} \sum_{i=1}^{k+1} \binom{k+1}{i} (-h_{N})^{i-1} \Big{\vert} +o(1)\Big{\vert} \sum_{i=1}^{\ell+1} \binom{\ell+1}{i} o(1)^{i-1} \Big{\vert} \\ & \hspace{0.8cm}+o(1)h_{N} \Big{\vert} \sum_{i=1}^{k+1} \binom{k+1}{i} (-h_{N})^{i-1}  \sum_{i=1}^{\ell+1} \binom{\ell+1}{i} o(1)^{i-1} \Big{\vert}\\ 
    & \leq 3\max \{h_{N},o(1)\}2^{k+\ell+2}.
\end{align*}
\end{linenomath*}
It follows that
\begin{linenomath*}
\begin{align*}
    \sum_{j=0}^{a_{N}} \lvert p^{(N)}_{j,u}-p_{j}\vert& \leq \sum_{j=0}^{a_{N}}12\max\{h_{N},o(1)\}\sum_{k+\ell=j}\frac{1}{k!}\left(\dfrac{c^{2}}{2} \right)^{k}\frac{x^{\ell}}{\ell!}\exp \left(-\left(\dfrac{c^{2}}{2}+x\right) \right)2^{k+\ell}\\ 
    &\leq 12\max \{h_{N},o(1)\} \exp\left(\frac{c^{2}}{2}+x\right) \\ 
    & \rightarrow 0,
\end{align*}
\end{linenomath*}
which ends the proof.
\end{proof}

\section{Coupling from below with (truncated) Galton-\\Watson processes }
\label{The Lower Galton-Watson Process}

\subsection{Establishing invasion} \label{Section Establishing invasion I}

Consider again the setting of Theorem \ref{Theorem} $(ii)$. The next proposition gives a lower bound on the probability that the parasite population infects at least $N^{\alpha}$ hosts for $0<\alpha<\beta$. 

\begin{Prop}
\label{lower bound probability reaching N alpha}
Consider the setting of Theorem \ref{Theorem} (ii) and let $0<\alpha<\beta$. Then
\begin{linenomath*}
\begin{equation*}
    \liminf_{N \to \infty}\mP \left( \exists n \in \mathbb{N}_0: \overline{I}^{(N)}_{n} \geq N^{\alpha} \right) \geq \pi(c,x).
\end{equation*}
\end{linenomath*}
\end{Prop}

\begin{remark}
Proposition \ref{lower bound probability reaching N alpha} together with the results from Section \ref{The Upper Galton-Watson Process} yield
\begin{linenomath*}
\begin{equation}
    \lim_{N \to \infty} \mP \left( \exists n \in \mathbb{N}_0: \overline{I}^{(N)}_{n} \geq N^{\alpha}\right)=\pi(c,x).
\end{equation}
\end{linenomath*}
\end{remark}

The remainder of this subsection is devoted to the proof of Proposition \ref{lower bound probability reaching N alpha}, which is given at the end of this subsection. First we introduce a simpler host-parasite model, see Definition \ref{def model with pairs of parasites}, that lower bounds the number of infected and removed hosts $\overline{I}^{(N)}$ of the original host-parasite model a.s. In this model hosts can get infected only by pairs of parasites moving along the same edge or by successful single parasites. In the following, we will refer to either a pair of parasites moving along the same edge or a successful single parasite as an \textit{infecting unit}.  We show then that whp the simpler process can be coupled with a Galton-Watson process from below until $N^\alpha$ hosts get infected, see Proposition \ref{coupling forward in time lower galton watson}. The total size of this lower Galton-Watson process reaches any level $\ell_N$ where $\ell_N \rightarrow \infty$ with asymptotic probability $\pi(c,x)$, see Lemma \ref{Lower GWP reaching large size then surviving whp}, in particular the level  $N^\alpha$. This yields the claimed lower bound.

\begin{Def}{(A simpler model involving only infecting units)} 
\label{def model with pairs of parasites}
\\
For a sequence $(N,d_{N},v_{N}, \rho_{N})_{N \in \N}$ introduce the following host-parasite model defined on the same random configuration model (with $N$ vertices and $d_N$ half-edges per vertex) as the original model.  Initially on each vertex a single host is placed. 
We start the infection process by infecting a randomly chosen host. A random number of infecting units is generated according to the following distribution with probability weights $(p^{(N)}_{j})_{j \in \mathbb{N}_{0}}$ where for all $   1\leq j\leq v_N$
\begin{linenomath*}
\begin{align}
\label{probability distribution number of pairs and alone phage}
        p^{(N)}_{j}:=\sum_{k+\ell=j, k\leq \floor{v_{N}/2}}w_k^{(N)} \cdot \binom{v_{N}-2k}{\ell}\rho_{N}^{\ell} \left(1-\rho_{N} \right)^{v_{N}-(2k+\ell)} ,
\end{align}
\end{linenomath*}
and 
\begin{linenomath*}
\begin{equation}
    p^{(N)}_{0}:=1-\sum_{j= 1}^{v_N} p^{(N)}_{k},
\end{equation}
\end{linenomath*}
where $w_k^{(N)}$ denotes the probability defined in \eqref{probability w_k}.
Afterwards, the host dies and the infection process  continues in discrete generations as follows. At the beginning of each generation, infecting units move, independently of each other, to nearest neighbour vertices along different, randomly chosen edges. If a host is attacked by at least one infecting unit, then the host gets infected. In each infected host, independently a random number of infecting units is produced according to the distribution $(p^{(N)}_{j})_{j \in \mathbb{N}_0}$. Afterwards the infected hosts and all the infecting units that infected the hosts die. If an infecting unit moves to an empty vertex, then it dies. \\ 
Denote by $J^{(N)}_{n}$ the number of hosts that get infected at generation $n$ in this simpler model and the epidemic process by $\mathcal{J}^{(N)}= (J^{(N)}_{n})_{n\in \mathbb{N}_0}$.
Furthermore we denote by $\overline{J}^{(N)}_n= \sum_{i=0}^n J_i^{(N)} $ the total number of hosts that got infected till generation $n$ in this simpler host-parasite model and by $\overline{\mathcal{J}}^{(N)} =(\overline{J}_n^{(N)})_{n\in \mathbb{N}_0} $ the corresponding process.
\end{Def}

\begin{Prop}\label{Prop coupling only pairs}
For all $N \in \N$ it is possible to couple ${\mathcal{J}}^{(N)}$ and $\mathcal{I}^{(N)}$ such that almost surely $\forall n \in \N_{0}$
\begin{linenomath*}
\begin{equation}
    \overline{J}^{(N)}_n  \leq  \overline{I}^{(N)}_{n}.
\end{equation}
\end{linenomath*}
\end{Prop}

\begin{proof}
Consider the same realisation of the configuration model for both host-parasite models and assume that the same host gets initially infected. \\
Enumerate the $d_N$ half-edges of each vertex and denote by $V^{(N)}_i \in \{0, \dots, v_N\}^{d_N}$ the occupancy vector of the half-edges linked to vertex $i$ (when host $i$ gets infected) by the $v_N$ offspring parasites generated at its infection in the original host-parasite model. By definition, the random variables  $(V^{(N)}_{i})_{1 \leq i \leq N}$ are iid. 
A coupling of $\overline{\mathcal{I}}^{(N)}$ and $\overline{\mathcal{J}}^{(N)}$  is obtained as follows. Use the same occupancy vector $V_i^{(N)}$ when host $i$ gets infected for the simpler host-parasite model but modify it as follows: Assume that in the original and in the simpler model the same single parasites are chosen to be successful and apply the subsequent rules:
\begin{itemize}
    \item If exactly $k$ pairs of parasites occupy $k$ different half-edges, the remaining parasites move separately along different half-edges, and if exactly $\ell$ of them are successful single parasites, for some $0 \leq k \leq \floor{v_{N}/2}$ and $0 \leq \ell \leq v_{N}$ such that $0 \leq 2k+\ell \leq v_{N}$, then in the simpler model all pairs of parasites and successful single parasites are kept and the remaining parasites are removed.
    \item If according to the occupancy vector $V_i^{(N)}$ at least one half-edge is occupied by at least three parasites, update $V_i^{(N)}$ for the simpler host-parasite model by removing all parasites, i.e. in particular no pairs of parasites or successful single parasite remain.
\end{itemize}
With this procedure the number of infecting units is distributed according to the distribution given in  \eqref{probability distribution number of pairs and alone phage}.
Moreover, hosts get either simultaneously infected in both host-parasite models or first in the original model and later possibly also in the simpler model. Hence, the number of infected hosts in the simpler model is bounded from above by $\overline{I}^{(N)}_{n}$ in any generation $n$.  
\end{proof}

Our next step is to couple $\mathcal{J}^{(N)}$  with the Galton-Watson process $\mathcal{Z}_{l}^{(N)}$ which is defined next.  

\begin{Def}(Lower Galton-Watson Process)
\label{lower branching process} \\
Let $0<\delta<\frac{1}{2}$ and $(a_N)_{N \in \N}$ be a sequence with $a_{N} \rightarrow \infty$ and $a_N \in o\left( \sqrt{d_N}\right)$. Furthermore assume $(\theta_N)_{N \in \N}$ is a $[0,1]$-valued sequence with
$\theta_{N} \rightarrow 0$. 
Let $\mathcal{Z}_{l}^{(N)}=\left( Z^{(N)}_{n,l}\right)_{n \in \N_0}$ be a Galton-Watson process with mixed binomial offspring distribution $\text{Bin}\left(\widetilde{Z}^{(N)},1-\theta_{N}\right)$, where the probability weights $\left(\widetilde{p}_k^{(N)}\right)_{k \in \N_0}$ of $\widetilde{Z}^{(N)}$ are
for all   $ 1 \leq j \leq a_{N}$
\begin{linenomath*}
\begin{align}
\label{first offspring distribution LGWP without removal 1}
    &\widetilde{p}_j^{(N)}:=\sum_{k+\ell=j}\left(\dfrac{(v_{N}-2a_{N})^{2}}{2d_{N}} \right)^{k}\frac{1}{k!}\exp \left(-\frac{v_{N}^{2}}{2d_{N}}\right) \left(1-\frac{1}{d_N^{\delta}} \right)\dfrac{((v_{N}-2a_{N})\rho_{N})^{\ell}}{\ell!}\left(1-\rho_{N}\right)^{v_{N}}, 
    \end{align}
    \end{linenomath*}
    and 
    \begin{linenomath*}
    \begin{align}\label{first offspring distribution LGWP without removal 2}
    &\widetilde{p}_0^{(N)}:=1-\sum_{j=1}^{a_N}\widetilde{p}_j^{(N)}.
\end{align}
\end{linenomath*}
Denote by $\Phi_{l}^{(N)}$ the generating function of the offspring distribution $\left(p_{k,l}^{(N)}\right)_{k \in \N_{0}}$
of $\mathcal{Z}_{l}^{(N)}$, and by $\pi_{l}^{(N)}$ and $q_{l}^{(N)}$ the survival and extinction probability of $\mathcal{Z}_{l}^{(N)}$, resp. Denote by $\overline{Z}^{(N)}_{n,l} := \sum_{i=0}^{n} Z_{i,l}^{(N)}$ the total size of the Galton-Watson process until generation $n$ and $\overline{\mathcal{Z}_{l}} = \left(\overline{Z}_{n,l}^{(N)}\right)_{n \in \mathbb{N}_0} $ the corresponding process.  
\end{Def}

\begin{Lem}
\label{Lower GWP reaching large size then surviving whp}
Let $(\ell_N)_{N \in \N}$ be a sequence with $\ell_N \rightarrow \infty $. Assume $Z_{0, l}^{(N)} =1$ a.s. Then
\begin{linenomath*}
\begin{equation*}
    \lim_{N \to \infty}\mP\left( \exists n \in \mathbb{N}_0: \overline{Z}_{n,l}^{(N)} \geq \ell_{N} \right)=\pi(c,x).
\end{equation*}
\end{linenomath*}
\end{Lem}

\begin{proof}

We proceed as in the proof of Proposition \ref{GWP reaching large size then surviving whp} and show that
\begin{linenomath*}
\begin{align}
\sum_{j=1}^{a_N} |\widetilde{p}_j^{(N)}- p_j| \rightarrow 0,
\end{align}
\end{linenomath*}
where $(p_j)_{j \in \mathbb{N}_0}$ denote the probability weights of the $\text{Pois}(c^2/2+x)$-distribution.
Using the same asymptotics as in the proof of Proposition \ref{GWP reaching large size then surviving whp}, we have for all $j\geq 1$
\begin{linenomath*}
\begin{equation*}
    \lvert \widetilde{p}^{(N)}_{j}-p_{j}\lvert \leq \sum_{k+\ell=j}\frac{1}{k!}\left(\dfrac{c^{2}}{2} \right)^{k}\frac{x^{\ell}}{\ell!}\exp \left(-\left(\dfrac{c^{2}}{2}+x\right) \right) \Big{\lvert} \left(1-h_{N}\right)^{k+1}(1+o(1))^{\ell+1}-1 \Big{\lvert},
\end{equation*}
\end{linenomath*}
where $(h_N)_N$ is an appropriate sequence with $h_{N}=\mathcal{O} \left(\frac{\max\{ a_{N},r_{N}\}}{\sqrt{d_{N}}} \right)$.
As in the proof of Proposition \ref{GWP reaching large size then surviving whp}, the last term can be upper bounded by
\begin{linenomath*}
\begin{align*}
    \vert \left(1-h_{N}\right)^{k+1}(1+o(1))^{\ell+1}-1 \vert \leq 3\max \{h_{N},o(1)\}2^{k+\ell+2}.
\end{align*}
\end{linenomath*}
It follows that
\begin{linenomath*}
\begin{align*}
    \sum_{j=1}^{a_{N}} \lvert \widetilde{p}^{(N)}_{j}-p_{j}\vert& \leq \sum_{j=1}^{a_{N}}12\max\{h_{N},o(1)\}\sum_{k+\ell=j}\frac{1}{k!}\left(\dfrac{c^{2}}{2} \right)^{k}\frac{x^{\ell}}{\ell!}\exp \left(-\left(\dfrac{c^{2}}{2}+x\right) \right)2^{k+\ell}\\ 
    &\leq 12\max \{h_{N},o(1)\} \exp\left(\frac{c^{2}}{2}+x\right) \rightarrow 0,
\end{align*}
\end{linenomath*}
which also implies that $\vert \widetilde{p}^{(N)}_{0}-p_0\vert \rightarrow 0$, because $a_N \rightarrow \infty$. Furthermore, we can estimate
\begin{linenomath*}
\begin{align*}
        \sum_{i=1}^{a_N} \lvert p^{(N)}_{i,l}-\widetilde{p}^{(N)}_{i} \lvert &\leq \sum_{i=1}^{a_N} \widetilde{p}^{(N)}_{i} \lvert 1-(1-\theta_{N})^{i} \lvert + \sum_{i=1}^{\infty} \sum_{j \geq i+1} \widetilde{p}^{(N)}_{j} \binom{j}{i} (1-\theta_{N})^{i}\theta_{N}^{j-i} \\
        & \leq \theta_{N} \sum_{i=1}^{\infty} \widetilde{p}^{(N)}_{i}2^{i}+\sum_{j=2}^{\infty}\widetilde{p}^{(N)}_{j} \sum_{i=1}^{j-1} \binom{j}{i} (1-\theta_{N})^{i}\theta_{N}^{j-i}\\ 
        & \leq \theta_{N} \left(\sum_{i=0}^{\infty} \frac{1}{i!}\left(\frac{c^{2}}{2}+x\right)^{i}\exp \left( \frac{-c^{2}}{2}+x\right) 2^{i}+\sum_{i=1}^{\infty} \lvert \widetilde{p}^{(N)}_{i}-p_{i} \lvert2^{i} \right) \\ & \quad \quad + \sum_{j=2}^{\infty} \widetilde{p}^{(N)}_{j} (1-(1-\theta_{N})^{j}) \\ 
        & \leq \theta_{N} \left[\exp \left( \frac{c^{2}}{2}+x\right)+12 \max\{h_{N},o(1) \} \exp \left(  \frac{c^{2}}{2}+x \right) \right]+\theta_{N}\sum_{j=2}^{\infty}\widetilde{p}^{(N)}_{j}2^{j}\\ 
        & \leq 2 \theta_{N} \left[\exp \left( \frac{c^{2}}{2}+x\right)+12 \max\{h_{N},o(1) \} \exp \left( \frac{c^{2}}{2}+x \right) \right] \\ 
        & \rightarrow 0,
\end{align*}
\end{linenomath*}
which implies $\vert p^{(N)}_{0,l}-\widetilde{p}_{0} \vert \rightarrow 0$ as well. An application of the triangle inequality ends the proof. 
\end{proof}

Next we show that the process $\overline{\mathcal{Z}}_{l}^{(N)}$ indeed bounds from below the number of infected hosts  $\overline{\mathcal{J}}^{(N)}$ in the simpler host-parasite model. Recall that $d_N \in \Theta(N^{\beta})$.

\begin{Prop}
\label{coupling forward in time lower galton watson}
Let $0<\alpha <\beta$, $\overline{\sigma}^{(N)}_{N^{\alpha}}:=\inf\{n \in \N_0:  \overline{J}^{(N)}_{n} \geq  N^{\alpha}\}$ and consider $\overline{\mathcal{Z}}_{l}^{(N)}$ with $\theta_N := \frac{2 N^\alpha \log(N)}{N - N^\alpha}$. Then
\begin{linenomath*}
\begin{equation*}
    \lim_{N \to \infty}\mP\Bigg{(} \overline{Z}_{n, l}^{(N)} \leq \overline{J}^{(N)}_n \ \forall n  < \overline{\sigma}^{(N)}_{N^{\alpha}} \Bigg{)}=1.
\end{equation*}
\end{linenomath*}
\end{Prop}

To prepare the proof of Proposition \ref{coupling forward in time lower galton watson} in the next lemma we estimate in the simpler host-parasite model the total number of infecting units $\overline{M}^{(N)}$ that can maximally be generated during the epidemic, and the total number of infecting units $M^{\alpha,(N)}$  that are generated until in total $N^\alpha$ hosts get infected.

\begin{Lem}
\label{upper bound number pairs of phages in graph with N/varphi(N) vertices}
Assume the conditions of Theorem \ref{Theorem} (ii) are fulfilled and $0<\alpha<1.$ Then we have
 \begin{linenomath*}
 \begin{align*}
    &\lim_{N \to \infty}\mP\left(\overline{M}^{(N)} \leq N \log(N) \right)=1, \\
    &\lim_{N \to \infty}\mP \left( M^{\alpha,(N)} \leq N^{\alpha} \log(N)\right)=1.
 \end{align*}
 \end{linenomath*}
\end{Lem}

\begin{proof}
Denote by $M_{i}^{(N)}$ the number of infecting units generated in host $i$ if it gets infected in the simpler model, i.e. $\overline{M}^{(N)} = \sum_{i=1}^{N} M_i^{(N)}$ and $M^{\alpha, (N)} \sim \sum_{i=1}^{N^\alpha} M_i^{(N)}$. By construction $M_{i}^{(N)}$ is distributed according to the probability distribution defined in \eqref{probability distribution number of pairs and alone phage} and the random variables $(M_{i}^{(N)})_{1 \leq i \leq N}$ are i.i.d. An application of Markov's inequality yields
\begin{linenomath*}
\begin{align*}
    & \mP \left( \sum_{i=1}^{N} M^{(N)}_{i} \geq N \log(N)\right) \leq \dfrac{\E[M^{(N)}_{1}]}{\log(N)}\rightarrow0, \\
    &\mP \left(\sum_{i=1}^{N^{\alpha}} M^{(N)}_{i}\geq N^{\alpha}\log(N) \right) \leq \dfrac{\E[M_{1}^{(N)}]}{\log(N)} \rightarrow 0, 
\end{align*}
\end{linenomath*}
because the expectations $\left( \E[M_{1}^{(N)}]\right)_{N \in \mathbbm{N}}$ are uniformly bounded. Indeed, recall the definition of the probability $w_k^{(N)}$ in \eqref{probability w_k}. We have
\begin{linenomath*}
\begin{align}
\label{mean number of pairs of phages is bounded}
    \E[M_{i}^{(N)}]&=\sum_{j=0}^{v_N} j \sum_{k+\ell=j, k\leq \floor{v_{N}/2}}w_k^{(N)}\cdot \binom{v_{N}-2k}{\ell}\rho_{N}^{\ell} \left(1-\rho_{N} \right)^{v_{N}-(2k+\ell)}  \notag\\ 
    & \leq \sum_{j=1}^{v_{N}}j \sum_{k+\ell=j} \left(\dfrac{v_{N}^2}{2d_{N}} \right)^{k} \frac{1}{k!}\cdot \frac{(v_{N}\rho_N)^{\ell}}{l!} \notag \\ 
    &=\sum_{j=1}^{v_N} j \frac{\left( \frac{v_N^{2}}{2d_N}+v_N\rho_N \right)^{j}}{j!} \notag \\ 
    & \leq \exp\left( \frac{v_N^{2}}{2d_N}+v_N\rho_N\right)\cdot \left(\frac{v_N^{2}}{2d_N}+v_N\rho_N\right) < \infty,
\end{align}
\end{linenomath*}
because $\frac{v_N^{2}}{2d_N} \rightarrow\frac{c^{2}}{2}$ and $v_N\rho_N \rightarrow x$. 
\end{proof}

\begin{proof}[Proof of Proposition \ref{coupling forward in time lower galton watson}]
Using the same kind of calculations as in the proof of
Proposition \ref{stochastic majoration pairs} we can show that for all $1 \leq j \leq a_{N}$, $\widetilde{p}^{(N)}_j \leq \mP(\bigcup_{k+\ell=j}D^{(N)}_{k,\ell})$, see Equations \eqref{k pairs and alone phages} and \eqref{asymptotic poisson (c,x) distribution}. In other words whenever a host gets infected we can estimate the number of infecting units, generated on the corresponding vertex according to the simpler model, from below by the corresponding number of offspring in the Galton-Watson process $\mathcal{Z}_{l}^{(N)}$, since $\widetilde{p}^{(N)}_{0}=1-\sum_{i=1}^{a_N}\Tilde{p}^{(N)}_j$.

However, in the host-parasite model 
 ``ghost'' infections may occur, when a) an already empty vertex is attacked by an infecting unit over a free half-edge, b) a vertex is attacked by more than one infecting unit or c) two infecting units attack an edge from different ends (and hence both infecting units hit empty vertices).

We will show next that each infecting unit generated before generation $\overline{\sigma}^{(N)}_{N^{\alpha}}$ is involved in one of the events a) or b) (independently of the other infecting units) with probability at most $\theta_N$. Furthermore, we will show that an event of type c) occurs before generation $\overline{\sigma}^{(N)}_{N^{\alpha}}$ only with negligible probability $o(1)$. Consequently, by removing infecting units with probability $\theta_N$ the number of offspring of infected hosts can whp be bounded from below by the number of offspring drawn according to the distribution with weights $(p_{k, l}^{(N)})_{k\in \mathbb{N}_0}$ from Definition \ref{lower branching process} for any generation $n<\overline{\sigma}^{(N)}_{N^{\alpha}}$. This yields the claimed coupling of $(\overline{J}_n^{(N)})_{n \in \mathbb{N}_0}$ and $(\overline{Z}^{(N)}_{n, l})_{n \in \mathbb{N}_0}$ before generation $\overline{\sigma}^{(N)}_{N^\alpha}$. 

We first control the probabilities of the events a) and b). 

a) Before generation $\overline{\sigma}^{(N)}_{N^{\alpha}}$ the number of free half-edges linked to an empty vertex is  bounded by $N^{\alpha}d_{N}$. Hence, the probability that an infecting unit on a half-edge gets connected to a half-edge of an empty vertex is bounded from above by $ \dfrac{N^{\alpha}d_{N}}{Nd_{N}-N^{\alpha}v_{N}} \sim \dfrac{1}{N^{1-\alpha}}$, since the total number of free half-edges is at least  $N d_N - N^\alpha v_N$.

b)
Before generation $\overline{\sigma}^{(N)}_{N^{\alpha}}$, the number of empty vertices in the graph is smaller than $N^{\alpha}$. Consequently, the probability that two infecting units attack the same vertex can be estimated from above by $\frac{d_{N}}{Nd_{N}-N^{\alpha}d_{N}} \sim \frac{1}{N}$. By Lemma \ref{upper bound number pairs of phages in graph with N/varphi(N) vertices} the total number of infecting units generated before generation  $\overline{\sigma}^{(N)}_{N^{\alpha}}$ is whp bounded by $N^{\alpha}\log(N)$. Hence, each infecting unit is involved in an event of type $b)$ with probability at most $N^{\alpha}\log(N) \cdot \dfrac{d_{N}}{Nd_{N}-N^{\alpha}d_{N}} \sim \dfrac{\log(N)}{N^{1-\alpha}}$. 

In summary, $\theta_N= 2 \cdot N^\alpha \log(N) \cdot \frac{d_N}{N d_N - N^\alpha d_N}$ yields an upper bound on the probability that an infecting unit is involved in one of the events of type a) or b). Since $\alpha < 1$ we have $\theta_N \in o(1)$.

It remains to show that whp events of type c) do not occur until generation $\overline{\sigma}_{N^\alpha}^{(N)}$.
According to Lemma \ref{upper bound number pairs of phages in graph with N/varphi(N) vertices} whp the number of infecting units that can be generated during the epidemic is at most $N\log(N)$  and before generation $\overline{\sigma}^{(N)}_{N^{\alpha}}$ the total number of generated infecting units can be estimated from above by $N^{\alpha}\log(N)$. Hence, whp we can estimate the probability that before time $\overline{\sigma}^{(N)}_{N^{\alpha}}$ none of the infecting units moves along an edge, on which end another infecting unit is located on, by 
\begin{linenomath*}
\begin{align*}
& \frac{Nd_{N}-N\log(N)}{Nd_{N}-1}\cdots \frac{Nd_{N}-N\log(N)-(N^{\alpha}\log(N)-1)}{Nd_{N}-1-2(N^{\alpha}\log(N)-1)} \\
     = & \frac{1}{\prod_{i=0}^{N^{\alpha}\log(N)-1}(Nd_{N}-1-2i)}\cdot \frac{(Nd_{N}-N\log(N))!}{(Nd_{N}-N\log(N)-N^{\alpha}\log(N))!} \\
     \geq & \left( \frac{(Nd_{N}-N\log(N)-N^{\alpha}\log(N))}{Nd_{N}} \right)^{N^{\alpha}\log(N)} \\ 
    = &\left(1-\frac{(N-N^{\alpha})\log(N)}{Nd_{N}} \right)^{N^{\alpha}\log(N)} \\
    =& 1- o(1),
\end{align*}
\end{linenomath*}
where
the last equality holds because $\alpha<\beta$.
\end{proof}

We conclude this section with the
proof of Proposition \ref{lower bound probability reaching N alpha}.

\begin{proof}[Proof of Proposition \ref{lower bound probability reaching N alpha}]
By Proposition \ref{Prop coupling only pairs} we can show the claim of the proposition for the event $\{\exists n \in \mathbb{N}_0: \overline{J}^{(N)}_n \geq N^{\alpha}\}$ instead of the event  $\{\exists n \in \mathbb{N}_0: \overline{I}^{(N)}_n \geq N^{\alpha}\}$. According to Proposition \ref{coupling forward in time lower galton watson} the process $\overline{\mathcal{J}}^{(N)}$ can whp be coupled from below by $\overline{\mathcal{Z}}_{l}^{(N)}$. By Lemma \ref{Lower GWP reaching large size then surviving whp}, the process $\overline{\mathcal{Z}}_{l}^{(N)}$ reaches at least the level $N^\alpha$ with asymptotic probability $\pi(c,x)$, which concludes the proof.
\end{proof}

\subsection{Growing further at exponential speed}
In Section \ref{Section Establishing invasion I} we showed that $N^\alpha$ hosts will get infected with asymptotic probability $\pi(c,x)$ for any $0<\alpha <\beta.$
 In Section \ref{Death of all bacteria in the graph} we will see that the total host population will go extinct whp in at most 2 generations if at least $N^{1-\frac{3}{4}\beta + 2\varepsilon}$ hosts get infected for any $\varepsilon >0$. 
If $\beta>\frac{4}{7}$ we have $1- \frac{3}{4} \beta < \beta$ and hence, with the results of the next section we can prove Theorem \ref{Theorem} (ii).
The aim of this section is to argue that also in the case $\beta \leq \frac{4}{7}$ whp  $N^{1-\frac{3}{4}\beta + 2\varepsilon}$ hosts will get infected once $N^\alpha$ hosts have been removed for some $0<\alpha < \beta$. Hence, we assume in the remainder of this subsection that $$\beta \leq \frac{4}{7}.$$ We will truncate the process  $\mathcal{Z}_{l}^{(N)}$ at certain time points. The resulting process $\mathcal{Z}_{t}^{(N)}=(Z^{(N)}_{n,t})_{n \in \N_{0}}$ grows asymptotically at the same speed as $\mathcal{Z}_{l}^{(N)}$  and can be coupled with $\overline{I}^{(N)}$ until the level $N^{1-\frac{3}{4}\beta+2\varepsilon}$ is reached. 
The coupling of $\mathcal{Z}_{l}^{(N)}$ with $\overline{I}^{(N)}$ fails if two infecting units attack an edge from two different ends at the same generation. In this case none of the two infecting units can reproduce because the vertices they are moving to are empty. Since in each generation, the number of infecting units involved in these events is small we can remove from time to time (the ancestors of) these infecting units without changing the asymptotic speed of exponential growth. 
Define $k_{0} \in \N$ through $k_0-1:=\sup \{k \in \mathbb{N}: k \beta \leq 1-\frac{3}{4}\beta\}$, in particular we have  
\begin{linenomath*}
\begin{equation}
    (k_{0}-1) \beta \leq 1-\frac{3}{4} \beta < k_{0}\beta.
\end{equation}
\end{linenomath*}

\begin{Def}\label{DefGraveProcess}
Let $\delta<\beta$ and $\varepsilon>0$ small enough such that $k_{0}(\beta-\delta)\geq 1-\frac{3}{4}\beta+2\varepsilon$. We define the process $\mathcal{Z}_{t}^{(N)}=(Z^{(N)}_{n,t})_{n \in \N_{0}}$ as follows. Assume $Z^{(N)}_{0,t}:=1$ almost surely, and let $\mathcal{Z}^{(N)}_{t}$ evolve as a Galton-Watson process with offspring distribution $(p^{(N)}_{k,l})_{k \in \N_{0}}$ until the level $N^{\beta-\delta}$ is reached, i.e. until time $\overline{\sigma}_{1}^{(N)}:=\inf \{ n \in \N_{0}: \overline{Z}^{(N)}_{n,t} \geq N^{\beta-\delta}\}.$  Set $Z^{(N)}_{\overline{\sigma}^{(N)}_{1}+1, t}:= \max\{Z^{(N)}_{\overline{\sigma}^{(N)}_{1},t}-N^{\beta-\frac{3}{2}\delta},0\}$. Assume that the process $Z^{(N)}_{t}$ is defined until generation $\overline{\sigma}_{i}^{(N)}+1$ for some $i \leq k_{0}-1$, then let the process evolve again as a GWP with offspring distribution $(p^{(N)}_{k,l})_{k \in \N_{0}}$ until generation $\overline{\sigma}_{i+1}^{(N)}:=\inf \{n \in \N_{0}: \overline{Z}^{(N)}_{n,t} \geq N^{(i+1)(\beta-\delta)}\}$ and set $Z^{(N)}_{\overline{\sigma}^{(N)}_{i+1}+1,t}:= \max\{Z^{(N)}_{\overline{\sigma}^{(N)}_{i+1},t}-N^{i\beta-\frac{2i+1}{2}\delta},0\}$.
\end{Def}

\begin{Prop}
\label{coupling truncated GWP and infected process}
Let $\overline{\tau}^{(N)}:=\inf \{n \in \N_{0}: \overline{I}^{(N)}_{n} \geq N^{1-\frac{3}{4}\beta+2\varepsilon
}\}$. Then
\begin{linenomath*}
\begin{equation*}
    \lim_{N \to \infty}\mP\left(\overline{Z}^{(N)}_{n,t} \leq \overline{I}^{(N)}_{n} \ \forall n \leq \overline{\tau}^{(N)}\right)=1.
\end{equation*}
\end{linenomath*}
\end{Prop}

\begin{proof}
The coupling of $\overline{Z}^{(N)}_{l}$ and $\overline{I}^{(N)}$ fails if 
two infecting units
attack an edge from both ends, because in this situation the corresponding branches in the Galton-Watson process have offspring but the corresponding infecting units do not infect any host. These infecting units cannot be treated independently and hence we cannot arrive at a coupling by thinning the Galton-Watson process. Instead we will remove the corresponding lines in the Galton-Watson process in pairs. 

If at some generation the number of infected hosts is $\mathcal{O}(N^{\alpha})$, then in this generation whp  $\mathcal{O}(N^{\alpha}\log(N))$ infecting units are generated, see Lemma \ref{upper bound number pairs of phages in graph with N/varphi(N) vertices}. Because whp the total number of infecting units is at most $N \log (N)$, see Lemma \ref{upper bound number pairs of phages in graph with N/varphi(N) vertices} again, an application of Chebyshev's Inequality yields that whp no more than $\mathcal{O}\left( \frac{N^{\alpha} \log(N) N \log(N)}{d_{N}N}\right)=\mathcal{O}\left( N^{\alpha-\beta} \log(N)^{2}\right)$ pairs of infecting units attack an edge from both ends.
Within the time intervals
$([\overline{\sigma}_{i}^{(N)} +1, \overline{\sigma}_{i+1}^{(N)}])_i$ 
in each generation
each individual has on average at least $c^2/2 + x  + o(1)$ offspring. Since within any time interval $[\overline{\sigma}_{i}^{(N)}+1, \overline{\sigma}_{i+1}^{(N)}]$ the process grows exponentially fast, for $1\leq  i \leq k_0$, whp at most $\mathcal{O}(N^{(i+1)(\beta-\delta)-\beta}\log^{3} (N))=o(N^{i\beta-\frac{2i+1}{2}\delta})$ pairs of infecting units are placed on two different ends of an edge. If we remove this number of pairs of infecting units at time $\overline{\sigma}_{i}^{(N)}+1$ and then let the process evolve like a GWP with offspring distribution $(p^{(N)}_{k,l})_{k \in \N_{0}}$, the total size of the resulting process whp lower bounds $\overline{I}^{(N)}$ until generation $\overline{\sigma}_{i+1}^{(N)}$. Continuing this algorithm till generation $\overline{\sigma}_{k_{0}}^{(N)}$, we arrive at the desired result. 
\end{proof}

\begin{Lem}
\label{Truncated Lower  GWP reaching large size then surviving whp}
Assume the process $\mathcal{Z}_{t}^{(N)}$ is constructed by means of the probability weights $(p_{k,l}^{(N)})_{k \in \mathbb{N}_0}$ with $\theta_N =\frac{2N^\alpha \log(N)}{N- N^\alpha} $ for some  $k_0 \beta <\alpha <1$. Assume $\varepsilon$ is small enough such that $1- \frac{3}{4} \beta + 2 \varepsilon < k_0 \beta$. Then
\begin{linenomath*}
\begin{equation*}
    \lim_{N \to \infty}\mP\left(\exists n \in \mathbb{N}_0: \overline{Z}_{n,t}^{(N)} \geq N^{1- \frac{3}{4} \beta + 2 \varepsilon} \right)=\pi(c,x).
\end{equation*}
\end{linenomath*}
\end{Lem}

\begin{proof}
Since $\mathcal{Z}_{t}^{(N)}$ and $\mathcal{Z}_{l}^{(N)}$ coincide until the level $N^\gamma$ is reached for any $\gamma \leq \beta - \delta$ an application of Lemma \ref{Lower GWP reaching large size then surviving whp} yields that the level $N^{\beta - \delta}$ is reached with asymptotic probability $\pi(c,x)$. If the level $N^{\beta - \delta}$ has been reached, the level $N^{1 - \frac{3}{4} \beta + 2 \varepsilon} \gg N^{\beta-\delta}$ will be reached whp. Indeed once a level $\ell_N$ has been reached by a supercritical GWP for some sequence $\ell_N \rightarrow \infty$, the GWP will explode whp. Since $Z^{(N)}_{\overline{\sigma}^{(N)}_{1}+1,t} \sim Z^{(N)}_{\overline{\sigma}^{(N)}_{1},t} =Z^{(N)}_{\overline{\sigma}^{(N)}_{1},\ell}$ and between generations $\overline{\sigma}^{(N)}_{1}+1$ and $\overline{\sigma}^{(N)}_{2}$, $\mathcal{Z}^{(N)}_t$ evolves as a supercritical GWP, we have $\overline{\sigma}^{(N)}_{2}<\infty$ whp. Repeating this argument $k_0-1$ times, we reach the level $N^{1-\frac{3}{4}\beta+2\varepsilon}$ whp. 
\end{proof}

From Proposition \ref{coupling truncated GWP and infected process} and Lemma \ref{Truncated Lower  GWP reaching large size then surviving whp} it follows that $\overline{\mathcal{I}}^{(N)}$ reaches the level $N^{1-\frac{3}{4}\beta+2\varepsilon}$ asymptotically with probability $\pi(c,x)$. Hence, for the proof of Theorem \ref{Theorem} (ii) it remains to show that $\overline{\mathcal{I}}^{(N)}$ reaches the level $N$ after hitting the level $N^{1-\frac{3}{4}\beta+2\varepsilon}$ whp. This is the topic of Section \ref{Death of all bacteria in the graph}.

\section{Final phase of the epidemic}
\label{Death of all bacteria in the graph}

In this section we consider again the setting of Theorem \ref{Theorem} (ii). We aim to show that once $N^\alpha$ hosts got infected eventually whp also the remaining hosts get infected.  
Assume in the following 
that $\varepsilon>0$ is small enough such that $1-\frac{3\beta}{4}+2\varepsilon<1-\frac{\beta}{2}$. Recall
\begin{linenomath*}
\begin{align*}
\overline{\tau}^{(N)} = \inf\{ n \in \mathbb{N}_{0}:  N^{1-\frac{3}{4} \beta +2 \varepsilon } \leq \overline{I}_n^{(N)} \},
\end{align*} 
\end{linenomath*}
and define
\begin{linenomath*}
\begin{align*}
\tau^{(N)} := \inf\{ n \in \mathbb{N}_{0}:  N^{1-\frac{3}{4} \beta +\varepsilon } \leq I_n^{(N)} \}.
\end{align*}\end{linenomath*}

\begin{Prop}
\label{death all hosts when beta>2/3}
For $\varepsilon$ defined as at the beginning of this section we have 
\begin{linenomath*}
\begin{equation*}
    \lim_{N \to \infty}\mP \left( \overline{I}^{(N)}_{\tau^{(N)}+2}=N \Big{\lvert} \overline{\tau}^{(N)} <\infty\right)=1.
\end{equation*}
\end{linenomath*}
\end{Prop}

The key observation for the proof of Proposition 5.1 is that infection by cooperation of parasites that attack a host from different edges determine the infection dynamics when $I_n^{(N)} \gg N^{1-\beta}$. Our assumptions on $\varepsilon$ guarantee that $\frac{(N^{1-\frac{3\beta}{4} + 2\epsilon} v_N)^2}{N} \ll N$. In Lemma \ref{ControlTgivenOverlineT} we will show that $\tau^{(N)} \leq  \overline{\tau}^{(N)}$ whp conditioned on  $\overline{\tau}^{(N)}< \infty$. Hence, we have    $N^{1- \frac{3}{4} \beta + \varepsilon}\leq  \overline{I}^{(N)}_{\tau^{(N)}} \ll N $ and 
 one generation further we have 
$   N^{1-\beta/2 +\varepsilon} \ll \overline{I}^{(N)}_{\tau^{(N)}+1} \leq N$ and also $N^{1-\beta/2 +\varepsilon} \ll {I}^{(N)}_{\tau^{(N)}+1}$. Consequently, in the following generation either the remaining hosts get infected, since $\frac{N^{1-\beta/2 +\varepsilon} v_N}{N} \gg N$ or (when already all hosts got infected) the number of removed hosts is $N$.

In the following we first state and prepare for the proof of Lemma \ref{ControlTgivenOverlineT}, then we  give the proof of this lemma and finish the section with the proof of Proposition \ref{death all hosts when beta>2/3}.

\begin{Lem}\label{ControlTgivenOverlineT}
For 
$\varepsilon$, $\tau^{(N)}$ as well as $\overline{\tau}^{(N)}$ defined as at the beginning of the section
\begin{linenomath*}
\begin{equation*}
    \lim_{N \to \infty}\mP \left( \tau^{(N)} \leq \overline{\tau}^{(N)} \Big{\lvert} \overline{\tau}^{(N)}<\infty \right)=1.
\end{equation*}
\end{linenomath*}
\end{Lem}

To prove Lemma \ref{ControlTgivenOverlineT} we control the time the approximating processes $(\mathcal{Z}_{t}^{(N)})_{N \in \N}$ need to reach some level $N^{\alpha}$.
We start with a rather classical result on branching processes. We give its proof in the Appendix for the sake of completeness.

\begin{Lem}
\label{almost sure convergence first hitting time N GWP}
Let $(Z_{n})_{n \in \N_{0}}$ be a Galton-Watson process with $m:=\E[Z_{1}]>1$. Consider the time $\tau_{N^{\alpha}}:=\inf \{ n \in \mathbb{N}_{0}: Z_{n} \geq N^{\alpha}\}$. Assume $Z_{0}=N^{\gamma}-\varphi(N)$ such that $Z_{0} \geq 1$, where $0\leq \gamma<\alpha$ and $\varphi(N) \in o(N^{\gamma})$. Denote by $W$ the almost sure limit of the non-negative martingale $(\frac{Z_{n}}{Z_{0}}m^{-n})_{n \in \N_{0}}$. \\ 
Conditioning on $\{W>0 \}$
\begin{linenomath*}
\begin{equation*}
    \dfrac{\tau_{N^{\alpha}}\log m}{(\alpha-\gamma)\log N} \rightarrow 1, \text{ almost surely}. 
\end{equation*}
\end{linenomath*}
\end{Lem}

Next we consider a family of Galton-Watson processes $((Z^{(\varepsilon)}_{n})_{n \in \N_{0}})_{\varepsilon>0}$, for which mean offspring numbers $m_\varepsilon$ are converging to some limit $m >1$ when $\varepsilon \downarrow 0$. In this case  the time to reach the level $N^\alpha$ from a level $N^{\gamma}$ is, conditioned on non-extinction, also not larger than $(1 + \delta)\frac{(\alpha-\gamma)\log N}{\log m}$  for $\varepsilon$ small enough and $\delta>0$.

\begin{Lem}
\label{upper bound in probability GWP reaching certain time}
Let $\mathcal{Z}^{(\varepsilon)}=(Z^{(\varepsilon)}_{n})_{n \in \N_{0}}$ be a Galton-Watson Process. Denote the mean number of offspring by $m_{\varepsilon}:=\E[Z^{(\varepsilon)}_{1}]=m-f(\varepsilon)$, where $f(\varepsilon)\underset{\varepsilon \to 0}{\longrightarrow}0$, and $m>1$. Introduce $W^{(\varepsilon)}$ the almost sure limit of the non-negative martingale $(Z^{(\varepsilon)}_{n}m_{\varepsilon}^{-1})_{n \in \N_{0}}$, and $\tau^{(\varepsilon)}_{N^{\alpha}}:=\inf \{n \in \mathbb{N}_{0}: Z^{(\varepsilon)}_{n} \geq N^{\alpha}\}$, the first time at which $\mathcal{Z}^{(\varepsilon)}$ reaches the size $N^{\alpha}$. \\
If $Z_{0}^{(\varepsilon)}=N^{\gamma}-\varphi(N)$ such that $Z_{0}^{(\varepsilon)} \geq 1$, where $0\leq \gamma<\alpha$ and $\varphi(N) \in o(N^{\gamma})$, then for all $\delta>0$ and for all $\varepsilon>0 \text{ small enough }$
\begin{linenomath*}
\begin{equation*}
    \lim_{N \to \infty}\mP \left(\tau^{(\varepsilon)}_{N^{\alpha}} \leq (1+\delta)\frac{(\alpha-\gamma)\log(N)}{\log(m)} \Big{\lvert} W^{(\varepsilon)}>0\right)=1.
\end{equation*}
\end{linenomath*}
\end{Lem}
\begin{proof}

Lemma \ref{almost sure convergence first hitting time N GWP} gives that for all $\delta>0$
\begin{linenomath*}
\begin{equation*}
    \lim_{N \to \infty} \mP \left( \tau^{(\varepsilon)}_{N^{\alpha}} \leq (1+\delta) \dfrac{(\alpha-\gamma)\log(N)}{\log(m_{\varepsilon})}\Big{\lvert} W^{(\varepsilon)}>0\right)=1.
\end{equation*}
\end{linenomath*}
And using that $m_{\varepsilon} \to m$ when $\varepsilon \to 0$, it directly follows the result of this Lemma. 
\end{proof}

Finally we consider a sequence of GWPes $( ( Z^{(N)}_{n})_{n \in \N_{0}})_{N \in \N}$, whose offspring distributions depend on $N$, and the level that we are interested to reach depends on $N$ as well.

\begin{Lem}
\label{upper bound time reaching certain size sequence GWP}
Let $\left( \left( Z^{(N)}_{n} \right)_{n \in \N_{0}} \right)_{N \in \N}$ be a sequence of GWPes whose offspring distributions are denoted by $(p_{k}^{(N)})_{k \in \mathbb{N}_{0}}$. Denote by $\Phi^{(N)}$ the corresponding sequences of generating functions of the offspring distributions. Consider the hitting times $\tau^{(N)}_{N^{\alpha}}:=\inf \{n \in \N_{0}: Z^{(N)}_{n} \geq N^{\alpha} \}$ and $\overline{\tau}^{(N)}_{N^{\alpha}}:=\inf \{n \in \N_{0}: \sum_{i=0}^{n}Z^{(N)}_{i} \geq N^{\alpha} \}$. Let $(p_{k})_{k \in \N_{0}}$ be a probability distribution and $\Phi$ its generating function, satisfying $1<m:=\Phi^{'}(1)<\infty$. Assume that
\begin{linenomath*}
\begin{equation}
\label{assumption}
    \sum_{k=0}^{\infty} \lvert p_{k}^{(N)} -p_{k} \lvert \rightarrow 0. 
\end{equation}
\end{linenomath*}
If $Z_{0}^{(N)}=N^{\gamma}-\varphi(N)$ such that $Z_{0}^{(N)} \geq 1$, where $0\leq \gamma<\alpha$ and $\varphi(N) \in o(N^{\gamma})$, then for all $\delta>0$
\begin{linenomath*}
\begin{equation*}
    \lim_{N \to \infty}\mP \left(\tau^{(N)}_{N^{\alpha}}\leq (1+\delta) \dfrac{(\alpha-\gamma) \log(N)}{\log m} \Big{\lvert} \overline{\tau}^{(N)}_{N^{\alpha}}<\infty\right)=1.
\end{equation*}
\end{linenomath*}
\end{Lem}

\begin{proof}
Using Assumption \eqref{assumption} it follows from the remark just before Lemma \ref{Lemma} that the sequence $(\Phi^{(N)})_{N \in \N}$ converges uniformly to the generating function $\Phi$. \\ 
Consider a family of natural numbers $(K_{\varepsilon})_{\varepsilon >0}$ satisfying $K_{\varepsilon}\underset{\varepsilon \to 0}{\longrightarrow}\infty$ and  $K_{\varepsilon}^{2}\varepsilon^{\gamma}\underset{\varepsilon \to 0}{\longrightarrow}0$, where $0<\gamma<1$. We introduce the GWP $\left(Z^{(\varepsilon)}_{n} \right)_{n \in \N_{0}}$, whose offspring distribution $\left(p_{k}^{(\varepsilon)}\right)_{k \in \N_{0}}$ is defined as follows. For all $1 \leq k \leq K_{\varepsilon}$
\begin{linenomath*}
\begin{align*}
        & p_{k}^{(\varepsilon)}:=\max \{p_{k}-\varepsilon^{\gamma},0\},
\end{align*}
\end{linenomath*}
and 
\begin{linenomath*}
\begin{align*}
    p_{0}^{(\varepsilon)}:=1-\sum_{k=1}^{K_{\varepsilon}}p_{k}^{(\varepsilon)}.
\end{align*}
\end{linenomath*}

This definition implies that the generating functions $\Phi^{(\varepsilon)}$ converge uniformly in $[0,1]$ to $\Phi$, as well as the mean number of offspring $m_{\varepsilon}:=\E[Z^{(\varepsilon)}_{1}]$ converges to $m$, when $\varepsilon \to 0$. Indeed, we have for all $0 \leq s \leq 1$
\begin{linenomath*}
\begin{align*}
    \lvert \Phi(s)-\Phi^{(\varepsilon)}(s) \lvert &\leq  \sum_{k=1}^{K_{\varepsilon}} s^{k}\varepsilon^{\gamma}+ \sum_{k=K_{\varepsilon}+1}^{\infty} s^{k}p_{k} +\left( p_{0}^{(\varepsilon)}-p_{0}\right) \\ 
    &\leq 2K_{\varepsilon}\varepsilon^{\gamma}+2\sum_{k=K_{\varepsilon}+1}^{\infty}p_{k} \\ 
    & \underset{\varepsilon \to 0}{\longrightarrow}0,
\end{align*}
\end{linenomath*}
since $K_{\varepsilon}\varepsilon^{\gamma} \underset{\varepsilon \to 0}{\longrightarrow}0$ and $K_{\varepsilon}\underset{\varepsilon \to 0}{\longrightarrow}\infty$. And also
\begin{linenomath*}
\begin{equation*}
    \lvert m-m_{\varepsilon} \lvert \leq \sum_{k=1}^{K_{\varepsilon}} k\varepsilon^{\gamma}+\sum_{k=K_{\varepsilon}+1}^{\infty}kp_{k} \leq K_{\varepsilon}^{2}\varepsilon^{\gamma}+\sum_{k=K_{\varepsilon}+1}^{\infty}kp_{k} \underset{\varepsilon \to 0}{\longrightarrow}0,
\end{equation*}
\end{linenomath*}
because $K_{\varepsilon}^{2}\varepsilon^{\gamma} \underset{\varepsilon \to 0}{\longrightarrow}0$ and $m<\infty$.\\ 
Moreover, Assumption \eqref{assumption} implies that $\sup_{k \in \N_{0}} \lvert p_{k}^{(N)}-p_{k}\lvert \rightarrow 0$, so there exists $N_{\varepsilon}$ such that  $N \geq N_{\varepsilon}$ and for all $k \in \N_{0}$
\begin{linenomath*}
\begin{equation*}
     p_{k}^{(N)} \geq \max\left\{p_{k}-\frac{\varepsilon^{\gamma}}{2}, 0\right\}.
\end{equation*}
\end{linenomath*}
Consequently, for all $N \geq N_{\varepsilon}$ we have $p_k^{(\varepsilon)} \leq p_k^{(N)}$ for all $ k \geq 1$ and $p_0^{(\varepsilon)} \geq p_0^{(N)}$. Hence, we can couple $\left( Z^{(\varepsilon)}_{n}\right)_{n \in \N_{0}}$ and $\left( Z^{(N)}_{n} \right)_{n \in \N_{0}}$ such that for all $n \in \N$
\begin{linenomath*}
\begin{equation*}
    Z^{(\varepsilon)}_{n} \leq Z^{(N)}_{n},
\end{equation*}
\end{linenomath*}
and 
\begin{linenomath*}
\begin{equation*}
    Z_{0}^{(N)}=Z_{0}^{(\varepsilon)}.
\end{equation*}
\end{linenomath*}
Lemma \ref{upper bound in probability GWP reaching certain time} and the convergence $m_{\varepsilon} \underset{\varepsilon \to 0}{\longrightarrow} m$ gives that for all $\delta >0$ and for all $\varepsilon>0 \text{ small enough}$
\begin{linenomath*}
\begin{equation*}
    \lim_{N \to \infty}\mP \left( \tau^{(\varepsilon)}_{N^{\alpha}} \leq (1+\delta)\frac{(\alpha-\gamma)\log (N)}{\log(m)}\Big{\lvert} W^{(\varepsilon)}>0 \right)=1, 
\end{equation*}
\end{linenomath*}
where $\tau^{(\varepsilon)}_{N^{\alpha}}:= \inf \{ n \in \N_{0}: Z^{(\varepsilon)}_{n} \geq N^{\alpha} \}$ and $W^{(\varepsilon)}$ is the almost sure limit of the non-negative martingale $\left(Z^{(\varepsilon)}_{n}/(Z^{(\varepsilon)}_{0}m_{\varepsilon}^{n})\right)_{n \in \N_0}$. The coupling yields 
\begin{linenomath*}
\begin{align*}
        1 & \geq \mP \left( \tau^{(N)}_{N^{\alpha}} \leq (1+\delta)\frac{(\alpha-\gamma)\log (N)}{\log(m)}\Big{\lvert} W^{(\varepsilon)}>0 \right) \\ &\geq \mP \left( \tau^{(\varepsilon)}_{N^{\alpha}} \leq (1+\delta)\frac{(\alpha-\gamma)\log (N)}{\log(m)}\Big{\lvert} W^{(\varepsilon)}>0 \right) \\ &\rightarrow 1,
\end{align*}
\end{linenomath*}
which yields that for all $\delta >0$ and for all $\varepsilon>0 \text{ small enough}$
\begin{linenomath*}
\begin{equation*}
    \lim_{N \to \infty}\mP \left( \tau^{(N)}_{N^{\alpha}} \leq (1+\delta)\frac{(\alpha-\gamma)\log (N)}{\log(m)}\Big{\lvert} W^{(\varepsilon)}>0 \right)=1.
\end{equation*}
\end{linenomath*}
Denote by $E_{N}:=\{ \tau^{(N)}_{N^{\alpha}} \leq (1+\delta)\frac{(\alpha-\gamma)\log (N)}{\log(m)}\}$, by $F_{\varepsilon}:=\{W^{(\varepsilon)}>0\}$, and by $G_{N}:=\{\overline{\tau}^{(N)}_{N^{\alpha}}<\infty\}$. The coupling implies that $F_{\varepsilon} \subset G_{N}$. Lemma \ref{Lemma} d) and the uniform convergence of the generating functions $\Phi^{(N)}$ to $\Phi$ give that $\lim_{N \to \infty} \mP (G_{N})=\pi$, where $\pi$ is the survival probability of the GWP with generating function $\Phi$. Lemma \ref{Lemma} a) and the uniform convergence of the generating functions $\Phi^{(\varepsilon)}$ give that $\lim_{\varepsilon \to 0} \mP(F_{\varepsilon})=\pi$. \\ 
We have
\begin{linenomath*}
\begin{align*}
        \mP (E_{N} \lvert F_{\varepsilon})&=\frac{\mP(E_{N} \cap F_{\varepsilon})}{\mP(F_{\varepsilon})}+\frac{\mP(E_{N}\cap (G_{N} \backslash F_{\varepsilon}))}{\mP(F_{\varepsilon})}-\frac{\mP(E_{N}\cap (G_{N} \backslash F_{\varepsilon}))}{\mP(F_{\varepsilon})} \\ 
        & \leq\mP(E_{N} \lvert G_{N})\cdot \frac{\mP(G_{N})}{\mP(F_{\varepsilon})},
\end{align*}
\end{linenomath*}
and taking the $\liminf_{N \to \infty}$ gives that 
\begin{linenomath*}
\begin{equation*}
    1 \leq \liminf_{N \to \infty}(\mP(E_{N} \lvert G_{N}) )\cdot\frac{\pi}{\mP(F_{\varepsilon})},
\end{equation*}
\end{linenomath*}
and finally by taking the limit when $\varepsilon \to 0$, we get
\begin{linenomath*}
\begin{equation*}
    \liminf_{N \to \infty} \mP(E_{N} \lvert G_{N})\geq 1,
\end{equation*}
\end{linenomath*}
and since it is a sequence of probability terms, it follows that 
\begin{linenomath*}
\begin{equation*}
    \lim_{N \to \infty} \mP(E_{N} \lvert G_{N})=1,
\end{equation*}
\end{linenomath*}
which is the result of this lemma.
\end{proof}

We apply the last lemma iteratively to the sequence of processes $(\mathcal{Z}_{t}^{(N)})_{N \in \N}$ introduced in Definition \ref{DefGraveProcess}.

\begin{Lem}
\label{upper bound random time total size LGWP reaches certain size}
Assume the process $\mathcal{Z}_{t}^{(N)}$ is constructed by means of the probability weights $\left(p_{k,l}^{(N)}\right)_{k \in \mathbb{N}_0}$ with $\theta_N =\frac{2N^\alpha \log(N)}{N- N^\alpha} $ for some  $k_0 \beta <\alpha <1$.
Then $ \forall \delta>0$
\begin{linenomath*}
\begin{equation*}
    \lim_{N \to \infty}\mP \left( \overline{\tau}^{(N)}_{N^{\alpha},t} \leq (1+\delta)\dfrac{\alpha\log(N)}{\log\left(\tfrac{c^2}{2}+x\right)} \Big{\lvert} \overline{\tau}^{(N)}_{N^{\alpha},t}<\infty\right)=1,
\end{equation*}
\end{linenomath*}
where $\overline{\tau}^{(N)}_{N^{\alpha},t}:=\inf \{n \in \N_{0}: \overline{Z}^{(N)}_{n,t} \geq N^{\alpha}\}$.
\end{Lem}

\begin{proof}
Since $\mathcal{Z}_{t}^{(N)}$ is, except at the time points $\overline{\sigma}_i^{(N)}$, a GWP, we can apply iteratively Lemma \ref{upper bound time reaching certain size sequence GWP} where Assumption \eqref{assumption} is obtained in the proof of Lemma \ref{Lower GWP reaching large size then surviving whp}. 
\end{proof}

Finally we come to the proof of Lemma \ref{ControlTgivenOverlineT}.

\begin{proof}[Proof of Lemma \ref{ControlTgivenOverlineT}]
If for every generation $n$ before $\overline{\tau}^{(N)}$, the number of infected hosts at generation $n$ satisfies $I^{(N)}_{n} \leq N^{1- \frac{3\beta}{4}+\varepsilon}$, then $\overline{\tau}^{(N)} \geq N^{\varepsilon}$. \\ 
But the coupling from below works whp at least until generation $\overline{\tau}^{(N)}$, and thanks to Lemma \ref{upper bound random time total size LGWP reaches certain size}, we know that the total size of the process $\mathcal{Z}_{t}^{(N)}$ will reach $N^{1- \frac{3\beta}{4}+2\varepsilon}$ within a time of order $\log(N)$. This implies that there exists $n \leq \overline{\tau}^{(N)}$ for which $I^{(N)}_{n} \geq N^{1- \frac{3\beta}{4}+\varepsilon}$. 
\end{proof}

\begin{Lem}\label{NotTooLarge}
For $\varepsilon$ defined as at the beginning of this section
\begin{linenomath*}
\begin{equation*}
    \lim_{N \to \infty} \mP \left( \overline{I}^{(N)}_{\overline{\tau}^{(N)}}\leq N^{1-\frac{\beta}{2}+5\varepsilon}\Big{\lvert} \overline{\tau}^{(N)}< \infty\right)=1.
\end{equation*}
\end{linenomath*}
\end{Lem}

\begin{proof}
The number of newly infected vertices is the sum of vertices that get attacked by successful single parasites or by several parasites simultaneously. The number of vertices that get infected by single successful parasites or pairs of parasites that move along the same edge denoted by $A^{(N)}$ is whp bounded from above by $N^{1-\frac{3}{4}\beta+3\varepsilon}$. We will show that the number of vertices that get infected by parasites attacking the vertex from different edges is whp bounded above by $N^{1-\frac{\beta}{2}+5\varepsilon}$. 

 At generation  $\overline{\tau}^{(N)}-1$ less than $N^{1- \frac{3\beta}{4}+2 \varepsilon}$ vertices are infected, and so there are less than $b_{N}:=v_{N}N^{1- \frac{3\beta}{4}+2\varepsilon}$ available parasites. Also the number of susceptible hosts is bigger than $N-N^{1- \frac{3\beta}{4}+2\varepsilon}$, and as we will show below whp they all have more than $d_{N}-\varphi(N)$ free half-edges for some sequence $(\varphi(N))_{N \in \N}$ where $\varphi(N)=O(1)$, see \eqref{Many free edges}.\\ 
 Denote by $D^{(N)}_{i}$ the number of free half-edges of vertex $i$ at generation $\overline{\tau}^{(N)}-1$. Assume we have $S_{N}$ boxes with box $i \leq S_{N}$ containing $D^{(N)}_{i}$ positions, and assume $b_{N}$ balls are distributed uniformly on the positions of the boxes, such that each position gets occupied at most once, and let $G^{(N)}_{i}$ be the number of balls put into box $i$. Then we have whp
 \begin{linenomath*}
 \begin{equation*}
     \overline{I}^{(N)}_{\overline{\tau}^{(N)}} \leq A^{(N)}+\sum_{i \in S_{N}} \1_{\{G^{(N)}_{i} \geq 2\}},
 \end{equation*}
 \end{linenomath*}
 because $A^{(N)}+\sum_{i \in S_{N}} \1_{\{G^{(N)}_{i} \geq 2\}} \geq N^{1-\frac{3}{4}\beta+2\varepsilon}$ whp. Denote by $G_{N}:=\sum_{i \in S_{N}}\1_{\{G^{(N)}_{i} \geq 2\}}$. We will show that 
 \begin{linenomath*}
 \begin{equation*}
     \lim_{N \to \infty}\mP \left( G_{N} \leq N^{1-\frac{\beta}{2}+5\varepsilon}\right)=1.
 \end{equation*}
 \end{linenomath*}
Denote by $T:=\sum_{j \in S_{N}}D^{(N)}_{j}$, $T_{i}:=T-D^{(N)}_{i}$ and $T_{i,j}:=T-\left(D^{(N)}_{i}+D^{(N)}_{j}\right)$. To estimate the expectation and variance of $G_{N}$ we estimate the probabilities of the events  $\{ G^{(N)}_{i} \leq 1\}$ and $\{ G^{(N)}_{i} \leq 1\} \cap \{ G^{(N)}_{j} \leq 1\}$ for $i \neq j$ conditioned on $S_{N}$.
Since $\mP \left( \{ N-N^{1-\frac{3\beta}{4}+2\varepsilon} \leq S_{N}\leq N \}\right)=1$ and $\mP \left(\bigcap_{i=1}^{S_{N}} \left\{d_{N}-\varphi(N) \leq D^{(N)} \leq d_{N}\right\} \Big{\lvert} S_{N}\right) \rightarrow 1$, Lemma \ref{Lemma appendix} can be applied whp. Hence, we have whp 
\begin{linenomath*}
\begin{align}\label{G < = 1}
        \mP \left(\{G^{(N)}_{i} \leq 1 \} \vert S_{N}\right)&=1-\frac{1}{2}\dfrac{b_{N}^{2}}{S_{N}^{2}}+\frac{1}{3}\dfrac{b^{3}_{N}}{S_{N}^{3}}-\dfrac{1}{8}\dfrac{b^{4}_{N}}{S_{N}^{4}}+\mathcal{O}\left(\dfrac{b^{5}_{N}}{S_{N}^{5}}\right),
\end{align}
\end{linenomath*}
and for all $i\neq j$

\begin{linenomath*}
\begin{equation*}
     \mP \left( \{ G_{i}^{(N)}\leq 1 \} \cap \{ G_{j}^{(N)}\leq 1 \} \lvert S_{N}\right)=1-\dfrac{b^{2}_{N}}{S_{N}^{2}}+\frac{2}{3}\dfrac{b^{3}_{N}}{S_{N}^{3}}+\mathcal{O}\left( \dfrac{b^{5}_{N}}{S_{N}^{5}}\right).
\end{equation*}
\end{linenomath*}
Using \eqref{G < = 1} we get
\begin{linenomath*}
\begin{equation*}
    \E[G^{(N)} \lvert S_{N}]= S_{N}\left(1-\mP\left( \{G_{i}^{(N)} \leq 1\}\vert S_{N}\right)\right)=\dfrac{b_{N}^{2}}{2S_{N}}+o \left( \dfrac{b_{N}^{2}}{S_{N}}\right),
\end{equation*}
\end{linenomath*}
and because $\mP \left( \{ N-N^{1-\frac{3\beta}{4}+2\varepsilon} \leq S_{N}\leq N \}\right)=1$, it follows that $E[G^{(N)}]=\mathcal{O}\left(N^{1-\frac{\beta}{2}+4 \varepsilon} \right)$.\\ 
The variance of $G^{(N)}$ conditioned on $S_N$ is estimated in Lemma \ref{Lemma appendix} as 
\begin{linenomath*}
\begin{align*}
    \V[G^{(N)} \lvert S_{N}]=\mathcal{O}\left( \frac{b_{N}^{4}}{S_{N}^{2}}\cdot \frac{b_{N}}{S_{N}}\right),
\end{align*}
\end{linenomath*}
as long as $S_N \sim N$.
The law of total variance yields
\begin{linenomath*}
\begin{align*}
    \V [G^{(N)}]&=\E [\V [G^{(N)} \lvert S_{N}]]+\V [\E[G^{(N)} \lvert S_{N}]] \\ 
    &=\mathcal{O} \left( \frac{b_{N}^{5}}{S_{N}^{3}}\right)+ \E[\E[G^{(N)}\lvert S_{N}]^{2}]-\E[G^{(N)}].
\end{align*}
\end{linenomath*}
The term $\E[\E[G^{(N)} \lvert S_{N}]^2]=\sum_{i=N-N^{1-\frac{3 \beta}{4}+2\varepsilon}}^{N} \mP(S_{N}=i) \left(\frac{b_{N}^{2}}{2i}+o\left( \frac{b_{N}^{2}}{i}\right) \right)^{2}\sim \frac{b_{N}^{4}}{N^{2}}$. This means that $\V [\E[G^{(N)} \lvert S_{N}]]$ can not exceed $\mathcal{O} \left(\frac{b_{N}^{4}}{N^{2}}\right)$, so an application of Chebyshev's Inequality yields the statement of the lemma. 

It remains to show that the number of free half-edges of each susceptible vertex is sufficiently close to $d_{N}$. Denote by $H^{(N)}_{i}$ the number of half-edges that are already formed for vertex $i$ in generation $ \overline{\tau}^{(N)}-1$, for $i \in \{1,\cdots,N\}$.
We show that
\begin{linenomath*}
\begin{equation}\label{Many free edges}
    \lim_{N \to \infty} \mP \left(H^{(N)}_{i} \leq \varphi(N), \forall i \in S^{(N)}_{\overline{\tau}^{(N)}-1}\right)=1,
\end{equation}
\end{linenomath*}
for any $\varphi(N)$ such that $\liminf_{N} \varphi(N) \geq 5$. Indeed, consider the following experiment: Assume we have $N-N^{1- \frac{3\beta}{4}+2\varepsilon}$ boxes, each with $d_{N}$ positions, and we distribute  uniformly at random $v_{N}N^{1- \frac{3\beta}{4}+2\varepsilon}$ balls  on the positions, such that each position gets occupied by at most one ball. Denote again by $(G^{(N)}_{i})_{i}$ the number of balls in box $i$. Then we have
\begin{linenomath*}
\begin{equation*}
    \mP \left(H^{(N)}_{i} \leq \varphi(N),\forall i \in S^{(N)}_{\overline{\tau}^{(N)}-1} \right) \geq \mP \left(G^{(N)}_{i} \leq \varphi(N), \forall i \leq N-N^{1- \frac{3\beta}{4}+2\varepsilon} \right),
\end{equation*}
\end{linenomath*}
\color{black}and assuming w.l.o.g. $\varphi(N)=5$, we have
\begin{linenomath*}
\begin{align*}
        \mP \left( \exists i: G^{(N)}_{i} \geq \varphi(N)\right)&=\dfrac{(N-N^{1- \frac{3\beta}{4}+2\varepsilon})\binom{v_{N}N^{1- \frac{3\beta}{4}+2\varepsilon}}{\varphi(N)}d_{N}!(d_{N}(N-N^{1- \frac{3\beta}{4}+2\varepsilon})-\varphi(N))!}{(d_{N}-\varphi(N))!(d_{N}(N-N^{1- \frac{3\beta}{4}+2\varepsilon}))!} \\ 
        & \leq N \cdot \frac{(v_{N}N^{1- \frac{3\beta}{4}+2\varepsilon})^{\varphi(N)}}{\varphi(N)!}\cdot\frac{d_{N}^{\varphi(N)}}{(d_{N}(N-N^{1- \frac{3\beta}{4}+2\varepsilon})-\varphi(N))^{\varphi(N)}} \\ 
        & \leq N \exp(\varphi(N))\left( \dfrac{v_{N}N^{1- \frac{3\beta}{4}+2\varepsilon}d_{N}}{(d_{N}(N-N^{1- \frac{3\beta}{4}+2\varepsilon})-\varphi(N))\varphi(N)}\right)^{\varphi(N)} \rightarrow 0.
\end{align*}
\end{linenomath*}
\end{proof}

\begin{Lem}\label{1-beta2}
In the setting of Theorem \ref{Theorem} (ii) there exists a constant $C>0$ such that
 \begin{linenomath*}
 \begin{equation*}
    \lim_{N \to \infty} \mP\left( I^{(N)}_{\tau^{(N)}+1}\geq C\cdot N^{1-\frac{\beta}{2}+\varepsilon} \Big{\lvert} \overline{\tau}^{(N)}< \infty\right)=1.
\end{equation*}
\end{linenomath*}
for $\varepsilon>0$ small enough.
\end{Lem}
\begin{proof}
According to Lemma \ref{ControlTgivenOverlineT}, $\tau^{(N)} \leq \overline{\tau}^{(N)}$ whp. Thus using Lemma \ref{NotTooLarge} the number of empty vertices at generation $\tau^{(N)}$ is whp at most $N^{1-\frac{\beta}{2}+5\varepsilon}$. By definition of $\tau^{(N)}$ there are at least $N^{1-\frac{3 \beta}{4}+\varepsilon}$ infected individuals, and so at least $\Theta(v_{N}N^{1-\frac{3 \beta}{4}+\varepsilon})$  parasites participate in new infections.\\ 
First we are going to show that the number of pairs of parasites present on infected vertices at generation $\tau^{(N)}$ are negligible compared to $v_{N}$. Denote by $A^{(N)}_{\tau^{(N)}}$ the number of parasites occupying an edge alone at generation $\tau^{(N)}$. Then for all functions $\varphi_{1}$, satisfying $\varphi_{1}(N) \rightarrow \infty$, we have
\begin{linenomath*}
\begin{align}
\label{NumberPairofParasites}
    \lim_{N \to \infty}\mP\left(A^{(N)}_{\tau^{(N)}} \geq N^{1-\frac{3\beta}{4}+\varepsilon}(v_{N}-\varphi_{1}(N)) \Big{\lvert} \overline{\tau}^{(N)}< \infty \right)=1.
\end{align}
\end{linenomath*}
Indeed, denote by $(K^{(N)}_{i})_{i\in \{1,...,N\}}$ the iid random variables giving the number of half-edges (connected to the vertices $i$, for $i\in \{1,...,N\}$) that are occupied by at least two parasites in the generations at which the vertices get infected. We have for all $0 \leq k \leq \lfloor{\frac{v_{N}}{2}}\rfloor$
\begin{linenomath*}
\begin{equation*}
    \mP(K^{(N)}_1=k)\leq  \dfrac{\binom{v_{N}}{2}\binom{v_{N}-2}{2}...\binom{v_{N}-2(k-1)}{2}}{k! \cdot d_N^{v_N}}.
\end{equation*}
\end{linenomath*}
Using Markov's inequality, we obtain that
\begin{linenomath*}
\begin{equation*}
    \mP\left( \sum_{i=1}^{N^{1-\frac{3 \beta}{4}+\varepsilon}} K^{(N)}_{i} \geq  N^{1-\frac{3 \beta}{4}+\varepsilon}\varphi_{1}(N) \right) \leq \dfrac{\E [K^{(N)}_1]}{\varphi_{1}(N)} \rightarrow 0,
\end{equation*}
\end{linenomath*}
since $\E(K^{(N)}_1)$ is uniformly bounded in $N$, see for a similar calculation Equation \eqref{mean number of pairs of phages is bounded}. \\ 
Denote by $H^{(N)}_{i}$ the number of half-edges that have already been formed for vertex $i$ till generation $\tau^{(N)}$. Using Lemma \ref{ControlTgivenOverlineT} and a similar computation as the one at the end of the proof of Lemma \ref{NotTooLarge} we obtain 
\begin{linenomath*}
\begin{equation}
\label{NumberAlreadyFormedEdges}
    \lim_{N \to \infty} \mP \left(H^{(N)}_{i} \leq \varphi_{2}(N) \ \forall i \in I^{(N)}_{{\tau}^{(N)}}\right)=1,
\end{equation}
\end{linenomath*}
where $\liminf_{N}\varphi_{2}(N) \geq 5$.\\ 
Thus, by \eqref{NumberPairofParasites} and \eqref{NumberAlreadyFormedEdges} the number of parasites that may cooperate by infecting a host from different edges is whp bounded from below by $N^{1-\frac{3\beta}{4}+\varepsilon}(v_{N}-2\widetilde{\varphi}(N))$ with $\widetilde{\varphi}(N):=\max \{\varphi_{1}(N),\varphi_{2}(N) \}$. \\ 
In addition it can also happen that a parasite attacks a half-edge on which another parasite is located. In this case, these two parasites cannot infect a host. An upper bound for the probability that a parasite is involved in such kind of event is whp $\frac{Nv_{N}}{Nd_{N}-2Nv_{N}}$. And so a lower bound on the number of available parasite is $N^{1-\frac{3 \beta}{4}+\varepsilon}(v_{N}-2\widetilde{\varphi}(N))(1-\frac{v_{N}}{d_{N}-2v_{N}})$. With this estimate we derive a whp lower bound on the number of infections occurring in the next generation.

Consider $N$ boxes, assume the $N^{1-\frac{\beta}{2}+5 \varepsilon}$ first boxes (corresponding to the empty vertices) contain $d_{N}$ positions and the remaining ones (corresponding to the susceptible vertices) contain  each$d_{N}-\varphi(N)$ positions and assume $\liminf_{N}\varphi(N)\geq 5$ as well as $\varphi(N)=o(d_{N})$. Distribute $ N^{1-\frac{3\beta}{4}+\varepsilon}(v_{N}-2\widetilde{\varphi}(N))(1-\frac{v_{N}}{d_{N}-2v_{N}})$ balls uniformly at random on the positions. Let $G^{(N)}$ be the number of boxes that contain $d_{N}-\varphi(N)$ positions and into which at least two balls are thrown. $G^{(N)}$ yields whp an estimate from below for the number of new infections. Using the same kind of computations as in the proof of Lemma \ref{NotTooLarge} (using Chebyshev's Inequality, estimating expectation and variance of $G^{(N)}$) we arrive at the statement of the lemma.
\end{proof}

\begin{Lem}\label{reachN}
Under the conditions of Theorem \ref{Theorem} (ii) it holds
\begin{linenomath*}
\begin{equation*}
    \lim_{N \to \infty} \mP \left(\overline{I}^{(N)}_{\tau^{(N)}+2} =N\Big{\lvert} 
    \overline{\tau}^{(N)}< \infty
    \right)=1.
\end{equation*}
\end{linenomath*}
\end{Lem}
\begin{proof}
 We aim to show that all hosts that have not been infected so far, get infected whp in generation $\tau^{(N)}+2$. 
 According to Lemma \ref{1-beta2} we have whp $I^{(N)}_{\tau^{(N)}+1}\geq C\cdot N^{1-\frac{\beta}{2}+2\varepsilon}$. Hence, we have whp at least $C\cdot N^{1-\frac{\beta}{2}+2\varepsilon} v_N$ parasites that may infect the remaining hosts. However, some of these parasites may be placed on already linked half-edges or occupy half-edges together with other parasites.  
 Hosts that got infected in generation $\tau^{(N)}+1$ have been attacked by at most one parasite in any generation $n \leq \tau^{(N)}$. By Lemma \ref{upper bound random time total size LGWP reaches certain size} whp $\tau^{(N)} \leq (1+\delta) \frac{\alpha \log N}{\log\left(\tfrac{c^2}{2}+x\right)}$ hence the number of formed edges is whp limited by $(1+\delta) \frac{\alpha \log N}{\log\left(\tfrac{c^2}{2}+x\right)}$ for any of these hosts for any $\delta>0$.
 Furthermore in generation $\tau^{(N)}$ we have according to Lemma \ref{NotTooLarge} and because $\tau^{(N)} \leq \overline{\tau}^{(N)}$ whp $ I_{\tau^{(N)}} \leq N^{1-\frac{\beta}{2} + 5 \varepsilon}.$ So by an application of Chebyshev's Inequality we can estimate that a host gets attacked in generation $\tau^{(N)} +1$ by at most $\frac{N^{1-\frac{\beta}{2} + 6 \varepsilon} v_N}{N} \sim N^{6\varepsilon}$ parasites with probability $1-\frac{1}{N^\varepsilon}$. 
 Consequently, at least a proportion $1- \frac{1}{N^\varepsilon}$ of the hosts infected at generation $\tau^{(N)}+1$ occupy whp a vertex with at least 
 \begin{linenomath*}
\begin{align*}
 e_N :=d_N - \left((1+\delta) \frac{\alpha \log N}{\log\left(\tfrac{c^2}{2}+x\right)} \right) -\frac{N^{1-\frac{\beta}{2} + 6 \varepsilon} v_N}{N}
 \end{align*}
 \end{linenomath*}
 free half-edges and the probability that the parasites generated in these hosts occupy a half-edge that has been linked before or that is occupied already by another parasite can be estimated from above by 
 \begin{linenomath*}
 \begin{align*}
     \frac{v_N + d_N - e_N}{d_N} \sim \frac{v_N}{d_N},
 \end{align*}
 \end{linenomath*}
 for $\varepsilon >0$ small enough.
 
 In summary, we have whp at least
\begin{linenomath*}
 \begin{align*}
 m_N:= C \cdot N^{1-\frac{\beta}{2} + 2 \varepsilon}\left(1- \frac{1}{N^{\varepsilon}}\right) v_N \left(1- \frac{v_N + d_N -e_N}{d_N}\right)
\end{align*} 
\end{linenomath*}
 free half-edges occupied with at least one parasite that may attack so far uninfected hosts.

Similarly an up to generation $\tau^{(N)} + 2$ uninfected host has whp at least
\begin{linenomath*}
\begin{align*}
    f_N:= d_N - \left((1+\delta) \frac{\alpha \log N}{\log\left(\tfrac{c^2}{2}+x\right)}\right) 
\end{align*}
\end{linenomath*}
 free half-edges.
So, the probability that an up to generation $\tau^{(N)} + 2$ uninfected host gets attacked by at most one of the $m_N$ parasites (and hence with high probability remains uninfected) can be estimated from above by
\begin{linenomath*}
 \begin{align*}
& \left(\left(1- \frac{f_N }{d_N N - v_N N }\right)^{m_N}   + \left(1-  \frac{f_N}{d_N N - v_N N }\right)^{m_N- 1} m_N \frac{f_N}{d_N N - v_N N } \right) (1+ o(1)) \\ & \quad \quad \sim N^{2\varepsilon} \exp(-N^{2\varepsilon}).
 \end{align*}
 \end{linenomath*}
 
The number of uninfected hosts at the beginning of generation $\tau^{(N)}+2$ is at most $N$. Consequently, the probability that at least one of these hosts remains uninfected till the end of generation $\tau^{(N)}+2$ can be estimated from above by a probability proportional to 
   \begin{linenomath*}
 \begin{align*}
      N \left(\exp(-N^{2\varepsilon}) N^{2\varepsilon}\right)   = o(1),
 \end{align*}
 \end{linenomath*}
 which yields the claim of Lemma \ref{reachN}.
\end{proof}

\begin{proof}[Proof of Proposition \ref{death all hosts when beta>2/3}]
According to Lemma \ref{ControlTgivenOverlineT} once $\overline{\mathcal{I}}_n^{(N)}$ has reached the level $N^{1-\frac{3\beta}{4} +2\varepsilon}$ also $\mathcal{I}^{(N)}$ has reached the level $N^{1-\frac{3\beta}{4} +\varepsilon}$. Moreover, according to Lemma \ref{NotTooLarge} whp $\overline{I}^{(N)}_{\overline{\tau}^{(N)}} \in \mathcal{O}(N^{1-\frac{\beta}{2} + 5 \varepsilon})$. Consequently, according to Lemma \ref{1-beta2}, the size of $\mathcal{I}^{(N)}$ is at generation $\tau^{(N)} +1$  whp at least $C \cdot N^{1-\frac{\beta}{2} + 2\varepsilon }$ for some appropriate constant $C>0.$ Finally, we can apply Lemma \ref{reachN}, which yields the result.
\end{proof}

\section{Proof of Theorem \ref{Theorem}}
\label{Proof of Main result}

\noindent
We start with the proof of Theorem \ref{Theorem} (ii): 

\noindent
For the upper bound on the invasion probability consider for a given $\ell_N>0$ 
the event 
\begin{linenomath*}
\begin{align*}
F_{\ell_N}^{(N)} := \{\exists n \in \mathbbm{N}_0: \overline{I}^{(N)}_n \geq \ell_N \}.    
\end{align*}
\end{linenomath*}
Then given $0< u\leq 1$ we have for any $\ell_N$ with $\ell_N \leq u N$ 
\begin{linenomath*}
\begin{align*}
 \mathbbm{P}(E_u^{(N)})\leq  \mathbbm{P}(F_{\ell_N}^{(N)}).
\end{align*}
\end{linenomath*}
For any sequence $(\ell_N)_{N\in \mathbbm{N}}$ with $\ell_N \rightarrow \infty$ and $\ell_N^3 v_N^2 \in o(N)$ we have by Proposition \ref{Coupling from above}
\begin{linenomath*}
\begin{align*}
\mathbbm{P}(F_{\ell_N}^{(N)}) \leq \mathbbm{P}\left( \exists n \in \mathbb{N}_0:  \overline{Z}^{(N)}_{n,u} \geq \ell_N\right),
\end{align*}
\end{linenomath*}
and by Proposition \ref{GWP reaching large size then surviving whp}
\begin{linenomath*}
\begin{align*}
\lim_{N \to \infty}  \mathbbm{P}\left( \exists n \in \mathbb{N}_0:  \overline{Z}^{(N)}_{n,u} \geq \ell_N\right)=\pi(c,x).
\end{align*}
\end{linenomath*}
Since for any given $0<u\leq 1 $ and any sequence $(\ell_N)$ with $\ell_N^3 v_N^2 \in o(N)$
we have for $N$ large enough $\ell_N \leq u N$. Hence, in summary 
\begin{linenomath*}
\begin{align*}
\limsup_{N \to \infty}\mathbbm{P}(E_u^{(N)})\leq \pi(c,x),
\end{align*}
\end{linenomath*}
which yields the claimed upper bound on the invasion probability.

For the lower bound we first apply Lemma \ref{Truncated Lower  GWP reaching large size then surviving whp}, which yields the lower bound $\pi(c,x) + o(1)$  on the probability that $N^\alpha$ hosts eventually get infected with $\alpha = 1-\frac{3}{4}\beta + 2 \varepsilon$.
Furthermore we can choose $\varepsilon >0$ small enough such that $\alpha < 1-\frac{\beta}{2}$. Then the assumptions of Proposition \ref{death all hosts when beta>2/3}
are fulfilled and we obtain the claimed upper bound on the invasion probability, since once the level $N^{1-\frac{3\beta}{4} + 2\varepsilon}$ is reached with probability 1+ o(1) the remaining hosts get infected as well, in particular any proportion $u$ of the host population for $0< u\leq 1$.   

\bigskip
\noindent
Proof of Theorem \ref{Theorem} (i): \\
\noindent
The proof of Theorem \ref{Theorem} (i) relies on the same arguments as the proof of Theorem \ref{Theorem} (ii).
Indeed, since $v_{N}=o(\sqrt{d_{N}})$ 
we have for any $c>0$ 
that the upper Galton-Watson process from Definition \ref{upper branching process}, where $v_{N}$ in this Definition is replaced by $c\sqrt{d_{N}}$, can be coupled with $\mathcal{I}^{(N)}$, such that  $\overline{\mathcal{I}}^{(N)}$ is bounded from above by $\overline{\mathcal{Z}}^{(N)}_u$ until $\overline{\mathcal{I}}^{(N)}$ is not further increasing or is exceeding the threshold $\ell_N$ for an appropriate sequence $(\ell_N)_{N\in \mathbbm{N}}$ fulfilling the conditions of Proposition \ref{Coupling from above}. Consequently, by Proposition \ref{GWP reaching large size then surviving whp} for all $0<u\leq 1$, the invasion probability satisfies $\mP(E^{(N)}_{u}) \leq \pi(c,x)+o(1)$. But since $x\leq 1$, we have $\lim_{c\downarrow 0} \pi(c,x)=0$ and so the statement follows, since $c>0$ was arbitrary. 
\bigskip

\noindent
Proof of Theorem \ref{Theorem} (iii):\\
\noindent
Trivially the invasion probability is upper bounded by 1. For the lower bound we can again rely on results of  the proof of Theorem \ref{Theorem} (ii). We consider, alongside the host-parasite model with the parameters $(d_N, v_N, \rho_N)$ fulfilling the conditions from Theorem \ref{Theorem} (iii), a host-parasite model with 
parameters $(d_N, v_N^{(c)}, \rho_N)$, where we set $v_N^{(c)}= c \sqrt{d_N}$, i.e. the parameters $(d_N, v_N^{(c)}, \rho_N)$ fulfill the conditions from Theorem \ref{Theorem} (ii). We couple these two host-parasite models by following, in the second host-parasite model, at each host infection instead of all $v_N$ parasite offspring only the first $v_N^{(c)}$ parasites. In this manner the process $\overline{\mathcal{I}}^{(N)}$ can be estimated from below by the corresponding process $\overline{\mathcal{I}}^{(N)}_c$ of the host-parasite model with parameters $(d_N, v_N^{(c)}, \rho_N)$. According to Theorem \ref{Theorem} (ii) a lower bound on the invasion probability of this model is $\pi(c,x) + o(1)$. Since for $c\rightarrow \infty$ we have $\pi(c,x) \rightarrow 1$ and $c$ can be chosen arbitrarily large this yields the claim of Theorem \ref{Theorem} (iii).\\
\\

\noindent
\textbf{Acknowledgements}\\

We thank Anton Wakolbinger for valuable hints and comments.
CP acknowledges support from the German Research Foundation through grant PO-2590/1-1.
We thank two referees for helpful comments on the manuscript.

\newpage

\begin{table}[htbp] 
    \centering
    \begin{tabular}{p{0.3 \linewidth}p{0.5 \linewidth}p{0.15 \linewidth}}
    \hline Notation & Meaning & Defined in \\ \hline 
    $d_{N}$ & number of edges per vertex \newline \setlength{\baselineskip}{12pt} scaling: $\Theta(N^{\beta}), 0<\beta<1$  & Section \ref{Model description and main results} \\ 
    \rule{0pt}{4ex}
    $v_N$ & number of offspring parasites, \newline \setlength{\baselineskip}{12pt} scaling in Theorem \ref{Theorem} (ii): $v_N\sim c \sqrt{d_N}$ & \rule[1.5mm]{7mm}{0.15mm} \texttt{"}  \rule[1.5mm]{7mm}{0.15mm} \\
    \rule{0pt}{4ex}
     $\rho_N$ & infection probability of a single parasite, \newline \setlength{\baselineskip}{12pt} scaling: $\rho_N v_N \rightarrow x \in [0,1]$ & \rule[1.5mm]{7mm}{0.15mm} \texttt{"}  \rule[1.5mm]{7mm}{0.15mm}  \\
     \rule{0pt}{4ex}
     $\pi(c,x)$ & survival probability of a GWP with \newline \setlength{\baselineskip}{12pt} Pois($\tfrac{c^2}{2}+x$) offspring distribution &  \\ 
     \rule{0pt}{4ex}
     $\mathcal{I^{(N)}}=\left( I^{(N)}_n\right)_{n \in \mathbb{N}}$ & process counting the number of infected hosts  &  \rule[1.5mm]{7mm}{0.15mm} \texttt{"}  \rule[1.5mm]{7mm}{0.15mm}  \\ \rule{0pt}{4ex}
     $\overline{\mathcal{I}}^{(N)}= \left( \overline{I}^{(N)}_n \right)_{n \in \mathbb{N}}$ & process counting the total number of hosts \newline \setlength{\baselineskip}{12pt} infected before generation $n$  & \rule[1.5mm]{7mm}{0.15mm} \texttt{"}  \rule[1.5mm]{7mm}{0.15mm}  \\ \rule{0pt}{4ex} 
     $\mathcal{Z}^{(N)}_{a}=\left( Z^{(N)}_{n,a}\right)_{n \in \mathbb{N}}$,  $a=u,l$ &  GWP used for approximating $\overline{I}^{(N)}$ \newline \setlength{\baselineskip}{12pt} from above (a=u) and from below (a=$l$) & Def.~\ref{upper branching process} and \newline \setlength{\baselineskip}{12pt}  Def.~\ref{lower branching process}, resp. \\ \rule{0pt}{4ex} 
     $\overline{\mathcal{Z}}^{(N)}_{a}=\left( \overline{Z}^{(N)}_{n,a}\right)_{n \in \mathbb{N}}$,  $a=u,l$ & total size of the process $\mathcal{Z}^{(N)}_a$ until generation $n$ &   \rule[1.5mm]{9mm}{0.15mm} \texttt{"}  \rule[1.5mm]{9mm}{0.15mm}  \\ \rule{0pt}{4ex}
     $\left( p_{k,a}^{(N)}\right)_{k \in \mathbb{N}_0}, a=u,l$ & probability weights of the offspring distribution of $\mathcal{Z}^{(N)}_a$ &  \rule[1.5mm]{9mm}{0.15mm} \texttt{"}  \rule[1.5mm]{9mm}{0.15mm}  \\ \rule{0pt}{4ex}
     $\mathcal{J}^{(N)}=\left( J^{(N)}_n\right)_{n \in \mathbb{N}}$ & process counting the number of infected hosts \newline \setlength{\baselineskip}{12pt} in the model from Definition \ref{def model with pairs of parasites} & Def. \ref{def model with pairs of parasites} \\  \rule{0pt}{4ex}
     $\overline{\mathcal{J}}^{(N)}=\left( \overline{J}^{(N)}_n\right)_{n \in \mathbb{N}}$ & process counting the total number of hosts \newline \setlength{\baselineskip}{12pt} infected before generation $n$ in the model from Definition \ref{def model with pairs of parasites} & \rule[1.5mm]{5mm}{0.15mm} \texttt{"}  \rule[1.5mm]{5mm}{0.15mm} 
    \end{tabular}
    \caption{Table of frequently used notation}
    \label{Table}
\end{table}

\bibliography{mybib}
\bibliographystyle{apalike}

\newpage
\begin{appendices}
\section{}
\renewcommand{\thesection}{\Alph{section}}

\begin{proof}[Proof of Lemma \ref{almost sure convergence first hitting time N GWP}]

Using the almost sure convergence of $\left(\frac{Z_{n}}{Z_0}m^{-n}\right)_{n \in \N_{0}}$ to $W$, it follows that 
for all  $\omega \in \{W>0 \}$, for all $\varepsilon>0$ there exists $\Tilde{n} \in \N_{0}$, such that for all  $n \geq \Tilde{n}$
\begin{linenomath*}
\begin{equation*}
(W-\varepsilon)m^{n} \leq \dfrac{Z_{n}}{Z_{0}} \leq (W+\varepsilon)m^{n}.
\end{equation*}
\end{linenomath*}
Introduce
\begin{linenomath*}
\begin{align*}
    &\tau_{\underline{N}^{\alpha}}:=\inf \left\{n \in \mathbb{N}_{0}: (W+\varepsilon)m^{n} \geq \frac{N^{\alpha} }{Z_{0}}\right\}, \\
    &\tau_{\overline{N}^{\alpha}}:=\inf \left\{n \in \mathbb{N}_{0}: (W-\varepsilon)m^{n} \geq \frac{N^{\alpha}}{Z_{0}} \right\}.
\end{align*}
\end{linenomath*}
We have $\tau_{\underline{N}^{\alpha}} \leq \tau_{N^{\alpha}} \leq \tau_{\overline{N}^{\alpha}}$, for $N$ large enough, and the following lower and upper bounds for $\tau_{\underline{N}^{\alpha}}$ and $\tau_{\overline{N}^{\alpha}}$ respectively hold for $\varepsilon$ small enough
\begin{linenomath*}
\begin{align*}
        &\tau_{\underline{N}^{\alpha}} \geq \dfrac{(\alpha-\gamma)\log N}{\log m}-\dfrac{\log (W+\varepsilon)}{\log m}-\dfrac{\log\left(1-\frac{\varphi(N)}{N^{\gamma}}\right)}{\log(m)}, \\ 
        & \tau_{\overline{N}^{\alpha}} \leq \dfrac{(\alpha-\gamma)\log N}{\log m}-\dfrac{\log (W-\varepsilon)}{\log m}-\dfrac{\log\left(1-\frac{\varphi(N)}{N^{\gamma}}\right)}{\log(m)}+1,
\end{align*}
\end{linenomath*}
which finally yields the following inequality
\begin{linenomath*}
\begin{equation*}
    1-\dfrac{\log (W+\varepsilon)+\log\left(1-\frac{\varphi(N)}{N^{\gamma}}\right)}{(\alpha -\gamma)\log N} \leq \dfrac{\tau_{N^{\alpha}}\log(m)}{(\alpha-\gamma)\log N} \leq 1-\dfrac{\log (W-\varepsilon)+\log\left(1-\frac{\varphi(N)}{N^{\gamma}}\right)-\log(m)}{(\alpha-\gamma)\log N}.
\end{equation*}
\end{linenomath*}
Taking the limit $N \rightarrow \infty$ concludes the proof. 
\end{proof}

For the proof of Lemma \ref{NotTooLarge} we need in addition to Lemma \ref{almost sure convergence first hitting time N GWP}  estimates on the number of vertices that get attacked by at least two parasites. For this purpose we consider the following experiment.

Let $(S_{N})_{N \in \N}$, $(D_{i}^{(N)})_{1 \leq i \leq S_{N}, N \in \N}$ be deterministic sequences of integers with $S_{N} \sim N$ and $D_{i}^{(N)}=d_{N}+O(1)$. Assume we have $S_{N}$ boxes with box number $i$ having $D_{i}^{(N)}$ many positions, and assume $b_{N}:=v_{N}N^{1-\frac{3\beta}{4}+2\varepsilon} \in \Theta( N^{1-\frac{\beta}{4}+2\varepsilon})$ balls are uniformly distributed on the positions of the boxes, such that each position gets occupied at most once, for some $\varepsilon >0$ small enough that $1- \frac{\beta}{4} + 2 \varepsilon <1$. Denote by $G_{i}^{(N)}$ the number of balls in box number $i$, and by $G^{(N)}:=\sum_{i=1}^{S_{N}}\1_{\{G_{i}^{(N)} \geq 2\}}$ the number of boxes containing at least $2$ balls. The following statements on the random variables $G_{i}^{(N)}$ and $G^{(N)}$ we apply in the proof of Lemma \ref{NotTooLarge}.  

\begin{Lem}\label{Lemma appendix}
\begin{linenomath*}
\begin{align}
        &\mP \left( \{ G^{(N)}_{i} \leq 1\} \right)=1-\frac{b_{N}^{2}}{2S_{N}^{2}}+\frac{b_{N}^{3}}{3S_{N}^{3}}-\frac{b_{N}^{4}}{8S_{N}^{4}}+\mathcal{O} \left(\frac{b_{N}^{5}}{S_{N}^{5}} \right), \\ 
        &\mP \left( \{ G^{(N)}_{i} \leq 1\} \cap \{ G^{(N)}_{j} \leq 1\} \right)=1-\frac{b_{N}^{2}}{S_{N}^{2}}+\frac{2}{3}\frac{b_{N}^{3}}{S_{N}^{3}}+\mathcal{O}\left( \frac{b_{N}^{5}}{S_{N}^{5}} \right), \\ 
        &\V [G^{(N)}]=\mathcal{O}\left( \frac{b_{N}^{4}}{S_{N}^{2}}\cdot \frac{b_{N}}{S_{N}}\right).
\end{align}
\end{linenomath*}
\end{Lem}

\begin{proof}
During the computation, we are using the following asymptotic estimates $
    \frac{1}{d_{N}}=o \left( \frac{b_{N}^{4}}{S_{N}^{4}}\right),$ $\frac{1}{S_{N}}=o \left( \frac{b_{N}^{4}}{S_{N}^{4}}\right), \frac{1}{S_{N}d_{N}}=o \left( \frac{b_{N}^{5}}{S_{N}^{5}}\right), \frac{b_{N}}{S_{N}d_{N}}=o \left( \frac{b_{N}^{5}}{S_{N}^{5}}\right), \frac{b_{N}}{S_{N}^{2}}=o \left( \frac{b_{N}^{5}}{S_{N}^{5}}\right), \frac{b_{N}^{2}}{S_{N}^{2}d_{N}}=o \left( \frac{b_{N}^{5}}{S_{N}^{5}}\right).$

To prepare the proof of the three estimates we first expand a few typical factors that will arise in the calculations of the two probability terms.
Denote by $T:=\sum_{i=1}^{S_{N}} D_{i}^{(N)}$ the total number of positions. We expand 
\begin{linenomath*}
\begin{align*}
        \left( \frac{T-D_{i}^{(N)}-b_{N}}{T-b_{N}}\right)^{b_{N}}&= \exp \left\{b_{N} \log \left[ 1-\frac{1}{S_{N}}\cdot\left(1+o\left(\frac{b_{N}^{4}}{S_{N}^{4}} \right)\right)\right]\right\} \\ 
        &=\exp\left(-\frac{b_{N}}{S_{N}}\right)\left(1+o\left(\frac{b_{N}^{5}}{S_{N}^{5}} \right)\right) \\ 
        &=1-\frac{b_{N}}{S_{N}}+\frac{b_{N}^{2}}{2S_{N}^{2}}-\frac{b_{N}^{3}}{6S_{N}^{3}}+\frac{b_{N}^{4}}{24S_{N}^{4}}+\mathcal{O}\left(\frac{b_{N}^{5}}{S_{N}^{5}} \right).
\end{align*}
\end{linenomath*}

and similarly we have for $k \in \N$ such that $\frac{b_{N}^{k}}{T^{k-1}}=o \left( \frac{b_{N}^{5}}{S_{N}^{5}} \right)$
\begin{linenomath*}
\begin{align*}
     & \frac{\left( 1-\frac{b_{N}}{T}\right)^{T}}{\left(1-\frac{b_{N}}{T-D_{i}^{(N)}} \right)^{T-D_{i}^{(N)}}}\\   
     = &\exp \left\{ T \left[ \log (1-\tfrac{b_{N}}{T})-\log \left(1-\frac{b_{N}}{T}\left[1+\tfrac{1}{S_{N}}+o\left(\tfrac{1}{S_{N}d_{N}}\right) \right]\right)\right]\right\}\\ 
    & \hspace{2cm}\cdot\exp \left( D_{i}^{(N)} \log\left(1-\tfrac{b_{N}}{T}\left[1+\mathcal{O}\left( \tfrac{1}{S_{N}}\right)\right] \right) \right) \\ 
        = &\exp \left( T \left(-\tfrac{b_{N}}{T}-\cdots-\tfrac{b_{N}^{k}}{kT^{k}} \right) \right) \exp \left( -\tfrac{b_{N}}{S_{N}}+o\left( \tfrac{b_{N}^{5}}{S_{N}^{5}}\right)\right)\\ 
        & \hspace{1cm} \cdot \exp\left\{ T\left[ \tfrac{b_{N}}{T}\left( 1+\tfrac{1}{S_{N}}+o\left( \tfrac{1}{S_{N}d_{N}}\right)\right)+\cdots +\tfrac{b_{N}^{k}}{kT^{k}}\left( 1+\tfrac{1}{S_{N}}+o\left( \tfrac{1}{S_{N}d_{N}}\right)\right)+o\left(\tfrac{b_{N}^{k}}{T^{k}} \right) \right]\right\}\\ 
        = &\exp \left(-\tfrac{b_{N}}{S_{N}}+o \left(\tfrac{b_{N}^{5}}{S_{N}^{5}}\right)\right)\exp \left( \tfrac{b_{N}}{S_{N}}+o\left( \tfrac{b_{N}^{5}}{S_{N}^{5}}\right)\right)\\ 
        =& 1+o \left(\tfrac{b_{N}^{5}}{S_{N}^{5}}\right).
\end{align*}
\end{linenomath*}

Using the asymptotic expansion of the factorial and the two previous estimates we get
\begin{linenomath*}
\begin{align}
\label{asymptotic factorial 1}
    & \dfrac{(T-D_{i}^{(N)})!}{(T-D_{i}^{(N)}-b_{N})!}\cdot\dfrac{(T-b_{N})!}{T!} \notag \\ = &\frac{(T-D_{i}^{(N)})^{T-D_{i}^{(N)}}}{(T-D_{i}^{(N)}-b_{N})^{T-D_{i}^{(N)}-b_{N}}}\cdot \left(\frac{T-D_{i}^{(N)}}{T-D_{i}^{(N)}-b_{N}}\right)^{\frac{1}{2}} \notag \\ 
    & \hspace{3cm}\cdot \frac{(T-b_{N})^{T-b_{N}}}{T^{T}}\cdot \left(\frac{T-b_{N}}{T}\right)^{\frac{1}{2}} \left(1+o\left( \frac{b_{N}^{5}}{S_{N}^{5}}\right) \right) \notag \\ 
    &=\frac{\left( 1-\frac{b_{N}}{T}\right)^{T}}{\left(1-\frac{b_{N}}{T-D_{i}^{(N)}} \right)^{T-D_{i}^{(N)}}}\cdot \left( \frac{T-D_{i}^{(N)}-b_{N}}{T-b_{N}}\right)^{b_{N}} \left(1-\frac{b_{N}}{T-D_{i}^{(N)}}\right)^{-\frac{1}{2}} \notag \\ & \hspace{3cm} \cdot\left(1-\frac{b_{N}}{T} \right)^{\frac{1}{2}}  \left(1+o\left( \frac{b_{N}^{5}}{S_{N}^{5}} \right) \right) \notag \\ 
    &=1-\frac{b_{N}}{S_{N}}+\frac{b_{N}^{2}}{2S_{N}^{2}}-\frac{b_{N}^{3}}{6S_{N}^{3}}+\frac{b_{N}^{4}}{24S_{N}^{4}}+\mathcal{O}\left( \frac{b_{N}^{5}}{S_{N}^{5}}\right),
\end{align}
\end{linenomath*}
and with similar calculations
\begin{linenomath*}
\begin{align}
\label{asymptotic factorial 2}
   \dfrac{(T-D_{i}^{(N)}-D_{j}^{(N)})!}{(T-D_{i}^{(N)}-D_{j}^{(N)}-b_{N})!}\cdot  \dfrac{(T-b_{N})!}{T!}=1-\frac{2b_{N}}{S_{N}}+\frac{2b_{N}^{2}}{S_{N}^{2}}-\frac{4b_{N}^{3}}{3S_{N}^{3}}+\frac{2b_{N}^{4}}{3S_{N}^{4}}+\mathcal{O}\left( \frac{b_{N}^{5}}{S_{N}^{5}}\right).
\end{align}
\end{linenomath*}
 Now we are ready to estimate the two probabilities
 \begin{linenomath*}
 \begin{align*}
    & \mP(\{G_{i}^{(N)} \leq 1 \}) \\= & \dfrac{(T-D_{i})!}{((T-D_{i}-b_{N})!}\cdot\dfrac{(T-b_{N})!}{T!}+b_{N}D^{(N)}_{i}\cdot \dfrac{(T-D_{i})!}{(T-D_{i}-(b_{N}-1))!}\cdot\dfrac{(T-b_{N})!}{T!}\\
         =& \dfrac{(T-D_{i})!}{(T-D_{i}-b_{N})!}\cdot\dfrac{(T-b_{N})!}{T!}\left(1+ \frac{b_{N}D^{(N)}_{i}}{T-D_{i}-(b_{N}-1)}\right)\\
         = & \dfrac{(T-D_{i})!}{(T-D_{i}-b_{N})!}\cdot\dfrac{(T-b_{N})!}{T!} \left(1+\frac{b_{N}}{S_{N}}+o\left( \frac{b_{N}^{5}}{S_{N}^{5}} \right) \right) \\ 
         =& \left( 1-\frac{b_{N}}{S_{N}}+\frac{b_{N}^{2}}{2S_{N}^{2}}-\frac{b_{N}^{3}}{6S_{N}^{3}}+\frac{b_{N}^{4}}{24S_{N}^{4}}+\mathcal{O}\left( \frac{b_{N}^{5}}{S_{N}^{5}}\right) \right) \left(1+\frac{b_{N}}{S_{N}}+o\left( \frac{b_{N}^{5}}{S_{N}^{5}} \right) \right) \\ 
         = & 1-\frac{b_{N}^{2}}{2S_{N}^{2}}+\frac{b_{N}^{3}}{3S_{N}^{3}}-\frac{b_{N}^{4}}{8S_{N}^{4}}+\mathcal{O} \left(\frac{b_{N}^{5}}{S_{N}^{5}} \right).
 \end{align*}
 \end{linenomath*}
\begin{linenomath*}
 \begin{align*}
     &\mP(\{G_{i}^{(N)} \leq 1 \} \cap \{ G_{j}^{(N)} \leq 1 \}) \\ = &\dfrac{(T-b_{N})!}{T!}\cdot \dfrac{(T-D_{i}^{(N)}-D_{j}^{(N)})!}{(T-D_{i}^{(N)}-D_{j}^{(N)}-b_{N})!}\\ 
     &\hspace{1cm}+b_{N}D_{i}^{(N)}\dfrac{(T-b_{N})!}{T!}\cdot \dfrac{(T-D_{i}^{(N)}-D_{j}^{(N)})!}{(T-D_{i}^{(N)}-D_{j}^{(N)}-(b_{N}-1))!}\\ &\hspace{1cm}+b_{N}D_{j}^{(N)}\dfrac{(T-b_{N})!}{T!}\cdot \dfrac{(T-D_{i}^{(N)}-D_{j}^{(N)})!}{(T-D_{i}^{(N)}-D_{j}^{(N)}-(b_{N}-1))!}\\ &\hspace{1cm}+b_{N}(b_{N}-1)D_{i}^{(N)}D_{j}^{(N)}\dfrac{(T-b_{N})!}{T!}\cdot\dfrac{(T-D_{i}^{(N)}-D_{j}^{(N)})!}{(T-D_{i}^{(N)}-D_{j}^{(N)}-(b_{N}-2))!} \\
    &=\dfrac{(T-b_{N})!}{T!}\cdot \dfrac{(T-D_{i}^{(N)}-D_{j}^{(N)})!}{(T-D_{i}^{(N)}-D_{j}^{(N)}-b_{N})!}\cdot\Bigg{[}1+\frac{b_{N}D_{i}^{(N)}}{T-D_{i}^{(N)}-D_{j}^{(N)}-(b_{N}-1)}\\ 
    & \hspace{2cm}+\frac{b_{N}D_{j}^{(N)}}{T-D_{i}^{(N)}-D_{j}^{(N)}-(b_{N}-1)}\\ & \hspace{2cm} +\frac{b_{N}(b_{N}-1)D^{(N)}_{i}D_{j}^{(N)}}{(T-D_{i}^{(N)}-D_{j}^{(N)}-(b_{N}-1))(T-D_{i}^{(N)}-D_{j}^{(N)}-(b_{N}-2))} \Bigg{]} \\
     &=\dfrac{(T-b_{N})!}{T!}\cdot \dfrac{(T-D_{i}^{(N)}-D_{j}^{(N)})!}{(T-D_{i}^{(N)}-D_{j}^{(N)}-b_{N})!} \\ & \hspace{2cm}\cdot \Bigg{(}1+\frac{2b_{N}}{S_{N}}\left(1+o\left( \frac{b_{N}^{4}}{S_{N}^{4}}\right)\right)+\frac{b_{N}^{2}}{S_{N}^{2}}\left( 1+o\left(\frac{b_{N}^{3}}{N^{3}} \right)\right) \Bigg{)} 
     \end{align*}
\end{linenomath*}
\begin{linenomath*}
 \begin{align*}
     &=\left(1-\frac{2b_{N}}{S_{N}}+\frac{2b_{N}^{2}}{S_{N}^{2}}-\frac{4}{3}\frac{b_{N}^{3}}{S_{N}^{3}}+\frac{2}{3}\frac{b_{N}^{4}}{S_{N}^{4}}+\mathcal{O}\left( \frac{b_{N}^{5}}{S_{N}^{5}} \right) \right) \cdot  \left(1+\frac{2b_{N}}{S_{N}}+\frac{b_{N}^{2}}{S_{N}^{2}}+o\left( \frac{b_{N}^{5}}{S_{N}^{5}}\right)\right) \\ 
     &=1-\frac{b_{N}^{2}}{S_{N}^{2}}+\frac{2}{3}\frac{b_{N}^{3}}{S_{N}^{3}}+\mathcal{O}\left( \frac{b_{N}^{5}}{S_{N}^{5}} \right).
\end{align*}
\end{linenomath*}
The estimate on the variance is obtained using the two previous computations
\begin{linenomath*}
\begin{align*}
        &\V [G^{(N)}]=\V \left[\sum_{i=1}^{S_{N}}\1_{\{ i \geq 2\}}\right]=\E \left[\left( \sum_{i=1}^{S_{N}}\1_{\{ G_{i}^{(N)} \geq 2\}}\right)^2\right]-\left(\E\left[\sum_{i=1}^{S_{N}}\1_{\{ G_{i}^{(N)} \geq 2\}}\right]\right)^{2} \\ 
        &=\sum_{i \neq j} \E \left[\1_{\{G_{i}^{(N)} \geq 2\}}\1_{\{G_{j}^{(N)} \geq 2\}}\right]+\E\left[ \sum_{i=1}^{S_{N}}\1_{\{G_{i}^{(N)} \geq 2 \}}\right]-\left(\E\left[\sum_{i=1}^{S_{N}}\1_{\{ G_{i}^{(N)} \geq 2\}}\right]\right)^{2} \\ 
       &=S_{N}(S_{N}-1)\mP\left( \{G_{i}^{(N)}\geq 2 \} \cap \{G_{j}^{(N)} \geq 2 \}\right)+S_{N}\mP(\{G_{i}^{(N)} \geq 2 \})-S_{N}^{2} \mP(\{G_{i}^{(N)} \geq 2 \}^{2} \\ 
        &=S_{N}^{2} \left(\mP(\{G_{i}^{(N)} \geq 2 \} \cap \{G_{j}^{(N)} \geq 2 \} )-\mP(\{G_{i}^{(N)} \geq 2\})^{2}\right) \\ & \hspace{2cm}+S_{N} \left(\mP(\{ G_{i}^{(N)} \geq 2 \})-\mP(\{G_{i}^{(N)} \geq 2 \} \cap \{G_{j}^{(N)} \geq 2 \} ) \right) \\ 
        &=S_{N}^{2} (\mP(\{G_{i}^{(N)} \leq 1 \} \cap \{G_{j}^{(N)} \leq 1 \})-\mP(\{G_{i}^{(N)} \leq 1 \})^{2}) \\ & \hspace{2cm} +S_{N}(\mP(\{ G_{i}^{(N)} \leq 1 \})-\mP(\{G_{i}^{(N)} \leq 1 \} \cap \{G_{j}^{(N)} \leq 1 \})) \\ 
        &=S_{N}^{2}\left[1-\frac{b_{N}^{2}}{S_{N}^{2}}+\frac{2}{3}\frac{b_{N}^{3}}{S_{N}^{3}}+\mathcal{O}\left( \frac{b_{N}^{5}}{S_{N}^{5}} \right)-\left( 1-\frac{b_{N}^{2}}{2S_{N}^{2}}+\frac{b_{N}^{3}}{3S_{N}^{3}}-\frac{b_{N}^{4}}{8S_{N}^{4}}+\mathcal{O}\left( \frac{b_{N}^{5}}{S_{N}^{5}}\right)\right)^{2} \right] \\ 
        &+S_{N} \left[1+\mathcal{O} \left(\frac{b_{N}^{2}}{S_{N}^{2}}\right)-\left(1+\mathcal{O}\left(\frac{b_{N}^{2}}{S_{N}^{2}}\right)\right)\right] \\ 
        &=S_{N}^{2}\left[1-\frac{b_{N}^{2}}{S_{N}^{2}}+\frac{2}{3}\frac{b_{N}^{3}}{S_{N}^{3}}+\mathcal{O}\left( \frac{b_{N}^{5}}{S_{N}^{5}} \right)- \left(1-\frac{b_{N}^{2}}{S_{N}^{2}}+\frac{2}{3}\frac{b_{N}^{3}}{S_{N}^{3}}+\mathcal{O}\left( \frac{b_{N}^{5}}{S_{N}^{5}} \right) \right)\right]+\mathcal{O}\left( \frac{b_{N}^{2}}{S_{N}}\right) \\ 
        &=\mathcal{O}\left( \frac{b_{N}^{5}}{S_{N}^{3}}\right)+\mathcal{O} \left( \frac{b_{N}^{2}}{S_{N}}\right) \\ 
        &=\mathcal{O}\left( \frac{b_{N}^{4}}{S_{N}^{2}}\cdot \frac{b_{N}}{S_{N}}\right).
\end{align*}
\end{linenomath*}

\end{proof}

\end{appendices}

\end{document}